\documentclass[11pt]{article}
\usepackage{graphicx}
\usepackage{amssymb}
\usepackage{epstopdf}
\usepackage{amsthm}
\usepackage{amsmath}
\usepackage{mathrsfs}
\usepackage{fullpage}

\title{Causality in gravitational theories with second order equations of motion}

\newcommand{\GG}{{\cal G}}

\theoremstyle{definition}

\def\be{\begin{equation}}
\def\ee{\end{equation}}
\def\ba{\begin{eqnarray}}
\def\ea{\end{eqnarray}}

\author{Harvey S. Reall\\ \\ {\small Department of Applied Mathematics and Theoretical Physics, University of Cambridge}\\ {\small Wilberforce Road, Cambridge CB3 0WA, United Kingdom} \\ hsr1000@cam.ac.uk}

\begin{document}
\maketitle

\begin{abstract}
This paper considers diffeomorphism invariant theories of gravity coupled to matter, with second order equations of motion. This includes Einstein-Maxwell and Einstein-scalar field theory with (after field redefinitions) the most general parity-symmetric four-derivative effective field theory corrections.  
A gauge-invariant approach is used to study the characteristics associated to the physical degrees of freedom in an arbitrary background solution. The symmetries of the principal symbol arising from diffeomorphism invariance and the action principle are determined. For gravity coupled to a single scalar field (i.e. a Horndeski theory) it is shown that causality is governed by a characteristic polynomial of degree $6$ which factorises into a product of quadratic and quartic polynomials. The former is defined in terms of an ``effective metric" and is associated with a ``purely gravitational" polarisation, whereas the latter generically involves a mixture of gravitational and scalar field polarisations. The ``fastest" degrees of freedom are associated with the quartic polynomial, which defines a surface analogous to the Fresnel surface in crystal optics. In contrast with optics, this surface is generically non-singular except on certain surfaces in spacetime. It is shown that a Killing horizon is an example of such a surface. It is also shown that a Killing horizon satisfies the zeroth law of black hole mechanics. The characteristic polynomial defines a cone in the cotangent space and a dual cone in the tangent space. The latter is used to define basic notions of causality and to provide a definition of a dynamical black hole in these theories.
\end{abstract}

\section{Introduction}

We will consider theories of gravity in $d$ spacetime dimensions described by a metric tensor $g_{\mu\nu}$ coupled to matter fields $\phi_I$, $I= 1, \ldots, N$, with a diffeomorphism-invariant action of the form
\be
\label{action1}
 S = \frac{1}{16\pi G} \int d^d x \sqrt{|g|} L(g,\phi_I)
\ee
for some scalar Lagrangian $L$. We will restrict attention to the class of theories for which the equations of motion are second order. As well as Einstein gravity minimally coupled to conventional matter fields, this class encompasses more exotic theories such as Lovelock theories \cite{Lovelock1971} (vacuum gravity in $d>4$ dimensions) and Horndeski theories \cite{Horndeski1974} (gravity coupled to a scalar field in $d=4$ dimensions). 

This class of theories also includes some important examples motivated by effective field theory (EFT). Einstein gravity minimally coupled to matter has a Lagrangian whose terms involve up to $2$ derivatives of the fields. In EFT one adds to this all possible scalars involving higher derivatives of the fields. The terms with the fewest derivatives give the leading corrections to Einstein gravity. Remarkably, in several important cases, one can use field redefinitions to arrange that the leading higher derivative corrections still give rise to second order equations of motion. We will discuss three examples of this.

Our first example is vacuum gravity. The leading EFT corrections to the Einstein-Hilbert Lagrangian have 4 derivatives. Using a field redefinition one can eliminate 4-derivative terms written in terms of the Ricci tensor, and arrange that the only 4-derivative term is the ``Gauss-Bonnet" term. This is topological for $d=4$ but not for $d>4$. If we neglect terms with more than 4 derivatives then we obtain Einstein-Gauss-Bonnet (EGB) theory, which has second order equations of motion and therefore belongs to the above class of theories (it is a Lovelock theory). 

The second example is the EFT of gravity coupled to a scalar field in $d=4$ dimensions. The leading EFT corrections to the minimally coupled 2-derivative theory involve terms with 4 derivatives. Field redefinitions can be used to write the 4-derivative terms in various different forms \cite{Weinberg:2008hq}. If one assumes a parity symmetry then one of these forms is preferred because, after neglecting terms with more than 4 derivatives, it gives rise to second order equations of motion \cite{Solomon:2017nlh,Glavan:2017srd,Kovacs:2020pns} (it is a Horndeski theory, described in section \ref{sec:EFT} below). Hence the above class of theories includes the leading EFT corrections to (parity-symmetric) Einstein-scalar theory in 4 dimensions. 

A third example is $d=4$ Einstein-Maxwell theory. In this case, after field redefinitions, one can reduce the possible parity-symmetric 4-derivative terms involving just the Maxwell tensor to $(F_{\mu\nu} F^{\mu\nu})^2$ and $(F_{\mu\nu} \tilde{F}^{\mu\nu})^2$ where $\tilde{F}_{\mu\nu}$ is the dual Maxwell tensor. These terms give second order equations of motion. The 4-derivative terms involving just the metric can be reduced to a topological term as for vacuum gravity. This leaves a possible 4-derivative interaction of the form $RFF$ where $R$ denotes the Riemann tensor. One can again exploit field redefinitions to write this term in the form $R_{\mu\nu\rho\sigma} \tilde{F}^{\mu\nu} \tilde{F}^{\rho\sigma}$, which gives second order equations of motion \cite{doi:10.1063/1.522837}. Thus the above class of theories includes Einstein-Maxwell theory with the leading parity symmetric 4-derivative EFT corrections. 

A minimal condition for a theory of the above type to ``make sense" classically is that it should admit a well-posed initial value problem. Recently it has been shown that this is indeed the case for Lovelock and Horndeski theories at ``weak coupling" \cite{Kovacs:2020pns,Kovacs:2020ywu}. The latter condition means roughly that the contribution of higher derivative terms to the equations of motion is small compared to the 2-derivative terms. Note that this condition is also required for validity of EFT. 

It is interesting to ask whether any of the important theorems of General Relativity can be extended to these theories. The first step in attempting to do this is to understand causal properties of these theories, which is the subject of this paper. 
The basic notion in the study of causality is the idea of a {\it characteristic hypersurface}. For example, the wavefront arising from a compactly supported perturbation is a characteristic hypersurface. In geometric optics, surfaces of constant phase are characteristic hypersurfaces. Characteristic hypersurfaces are defined in a general background solution as follows. Consider the equations of motion linearized around the background solution. These take the form
\ba
\label{lin_eqs}
 P_{gg}^{\mu\nu\rho\sigma\alpha\beta} \partial_\alpha \partial_\beta \delta g_{\rho\sigma} + P_{gm}^{\mu\nu I\alpha \beta}\partial_\alpha \partial_\beta \delta \phi_I + \ldots &=& 0 \nonumber \\ P_{mg}^{I \mu \nu\alpha \beta} \partial_\alpha \partial_\beta \delta g_{\mu\nu} + P_{mm}^{IJ\alpha\beta} \partial_\alpha \partial_\beta \delta \phi_J  + \ldots &=&0
 \ea
where the ellipses denotes terms with fewer than $2$ derivatives acting on $(\delta g_{\mu\nu}, \delta \phi_I)$. Subscripts ``g" and ``m" refer to ``gravity" or ``matter". The coefficients $P_{gg}$, $P_{gm}$, etc of the 2-derivative terms are tensors that can depend on the background fields $(g_{\mu\nu},\phi_I)$ and their first and second derivatives. These coefficients are assembled into a matrix called the {\it principal symbol} defined as follows. Let $\xi_\mu$ be an arbitrary covector. Then the principal symbol is
\be
\label{PS_def}
 {\cal P}(\xi) = \left( \begin{array}{cc} P_{gg}^{\mu\nu\rho\sigma\alpha\beta}\xi_\alpha \xi_\beta & P_{gm}^{\mu\nu I\alpha \beta}\xi_\alpha \xi_\beta \\ P_{mg}^{I \mu\nu\alpha \beta}\xi_\alpha \xi_\beta & P^{IJ\alpha \beta}_{mm}\xi_\alpha \xi_\beta \end{array} \right)
\ee
We can regard this matrix as acting on ``polarization vectors" of the form $T\equiv (t_{\mu\nu}, t_I)$ where $t_{\mu\nu}$ is symmetric. However, owing to the diffeomorphism invariance of our theory, it is better to regard it as acting on gauge equivalence classes of polarisations \cite{christodoulou2008mathematical}, a notion we review below. Such classes correspond to ``physical polarisations". One then defines a covector $\xi_\mu$ to be characteristic if there exists a non-zero equivalence class $T$ satisfying the characteristic equation ${\cal P}(\xi)T=0$. A hypersurface is characteristic iff its normal is a characteristic covector. Associated with any such hypersurface is a physical polarisation (or space of polarisations) that, in geometric optics, can propagate along that hypersurface. 

Causality is determined by algebraic properties of the principal symbol and so we must start by studying these algebraic properties. In section \ref{sec:princ}, I will show that the principal symbol must possess certain symmetries. These follow from the action principle and from diffeomorphism invariance. These symmetries are particularly restrictive in low spacetime dimensions. For $d=4$ they imply that the tensor $P_{gg}^{\mu\nu\rho\sigma\alpha\beta}$ can be written in terms of an ``effective metric": a symmetric tensor $C_{\mu\nu}$ depending on the background fields and their first and second derivatives. For a weakly coupled theory, $C_{\mu\nu}$ is close to $g_{\mu\nu}$. Next, in section \ref{sec:chars}, I will study the characteristic equation for a general theory, focusing on the $d=4$ case. The analysis splits into two cases. In the first case, $\xi_\mu$ is non-null w.r.t. $(C^{-1})^{\mu\nu}$ (the inverse of $C_{\mu\nu}$) and the gravitational polarization $t_{\mu\nu}$ is determined (up to gauge) by the matter polarization $t_I$. In the second case, $\xi_\mu$ is null w.r.t. $(C^{-1})^{\mu\nu}$. Whether or not this case arises reduces to the condition that a ``Weyl-like" tensor constructed from the principal symbol and $t_I$ should admit $\xi_\mu$ as a principal null direction. 

In section \ref{sec:horn}, I will consider $d=4$ theories of gravity coupled to a single scalar field, i.e., Horndeski theories. In this case I will show, for a general background solution, that $\xi_\mu$ is characteristic iff $p(\xi)=0$ where the {\it characteristic polynomial} is
\be
\label{p_fact}
 p(\xi) = (C^{-1})^{\mu\nu}\xi_\mu \xi_\nu Q(\xi)
\ee
where $Q(\xi)$ is a homogeneous quartic polynomial in $\xi$ with coefficients depending on the backgrounds fields and their first and second derivatives. Clearly $p(\xi)$ is a homogeneous polynomial of degree $6$ that factorises into the product of a quadratic and a quartic polynomial. For simple theories, or symmetrical backgrounds (e.g. a FLRW cosmology), $Q(\xi)$ also factorises, into a product of quadratic polynomials of the form $(C^{-1})^{\mu\nu} \xi_\mu \xi_\nu F^{\rho\sigma} \xi_\rho \xi_\sigma$, where $F^{\mu\nu}$ is another effective metric. However this factorisation does not occur for generic backgrounds of generic theories (such as the EFT of a scalar field coupled to gravity). 

This result implies that, in a Horndeski theory, the normal $\xi_\mu$ to a (physical) characteristic surface must satisfy either the quartic equation $Q(\xi)=0$ or the quadratic equation $(C^{-1})^{\mu\nu} \xi_\mu \xi_\nu=0$. The set of solutions $\xi_\mu$ of $p(\xi)=0$ defines the {\it characteristic cone} in the cotangent space at any point. So the result \eqref{p_fact} shows that the characteristic cone is the union of a quadratic cone and a quartic cone. 

There is a close similarity with the theory of electromagnetic waves in an anisotropic crystal \cite{born2013principles}, or elastic waves in an anisotropic solid \cite{duff1960cauchy}. In the former case, characteristics $\xi_\mu$ must satisfy a quartic equation and in the latter case they must satisfy an equation of degree $6$, as in our case. The analogy is particularly close for an elastic solid with hexagonal symmetry for which the characteristic polynomial factorises into a quadratic and quartic polynomial as in \eqref{p_fact} \cite{duff1960cauchy}. 

One can visualise the characteristic cone by taking a cross-section (of constant $\xi_0$ w.r.t. a suitable basis) to define a ``slowness surface" in $\mathbb{R}^3$. I will show that, for a weakly coupled theory, the slowness surface has three sheets, corresponding to the three physical degrees of freedom. This surface is the union of an ellipsoid defined by the quadratic equation with a $2$-sheeted surface defined by the quartic equation, with the quadratic ellipsoid lying between the sheets of the quartic surface. In optics (for a biaxial crystal), the analogous surface, sometimes called the Fresnel surface, has $4$ singular points where two sheets of the quartic meet. This gives rise to the phenomenon of conical refraction \cite{born2013principles}. In our case I will show that, for a generic background of a generic theory, the slowness surface surface has $4$ ``double points" at which the quadratic and quartic surfaces meet, however, generically they do so smoothly, which implies that there is no conical refraction. This is similar to the case of some hexagonally-symmetric materials in elasticity (e.g. Zinc \cite{musgrave1954propagation}). In our case, for a generic background of a generic theory, there will be special (non-generic) points in spacetime for which a double point is replaced by a ``triple point" where all three sheets of the slowness surface meet. I will argue that these special points fill out hypersurfaces in spacetime. 

I will give an expression for the physical polarisations associated with each sheet of the characteristic cone. The quadratic cone is associated with a ``purely gravitational" polarisation. In particular, this means that, in any (weakly coupled) background, for any Horndeski theory, there is always a physical graviton polarisation that decouples from the scalar field in the geometric optics limit (and propagates on the null cone of $C_{\mu\nu}$). However, in a generic theory, this is not the ``fastest" degree of freedom. The latter is associated with the inner sheet of the quartic cone. The polarisations associated with the sheets of the quartic cone generically involve mixing between the gravitational and scalar field degrees of freedom. 

Associated with the region enclosed by the inner sheet of the characteristic cone one can define a dual cone in the tangent space at any point. This dual cone governs causality in these theories, i.e., it provides the appropriate generalisation of the usual ``light cone" of GR. The dual cone can be used to generalise standard GR definitions to this class of theories. I will use it to provide a definition of the black hole region in an asymptotically flat spacetime. 

I will discuss in detail the case of a spacetime admitting a Killing horizon. It turns out that such a horizon is an example of a surface on which the slowness surface has a triple point. However, this triple point is of a special type which ensures that conical refraction does not occur and, within the horizon, causality reduces to the usual notion of causality w.r.t. the metric. I will also prove some results about the surface gravity of a Killing horizon. First, the surface gravity is constant if the theory is weakly coupled on the Killing horizon, i.e., the zeroth law of black hole mechanics holds. Second, the surface gravity defined w.r.t. the effective metric $C_{\mu\nu}$ is the same as that defined w.r.t. $g_{\mu\nu}$. 

This paper uses standard notions of causality from PDE theory. However, as discussed above, some of the theories discussed in this paper are motivated by EFT. The regime of validity of EFT does not include waves of arbitrarily short wavelength. I will explain why this means that one cannot use geometric optics to distinguish between  the characteristic cone defined by \eqref{p_fact} and the usual null cone of the metric.  

I will end in section \ref{sec:discuss} by discussing some possibilities for future research. 

\subsection*{Notation and conventions}

Lower case Greek indices are tensor indices. I consider theories involving a metric tensor $g_{\mu\nu}$ with positive signature. All index raising and lowering will be performed using this metric tensor (rather than the effective metric $C_{\mu\nu}$). My convention for the Riemann tensor is $R^\mu{}_{\nu\rho\sigma} = 2\partial_{[\rho} \Gamma^\mu_{|\nu| \sigma]}+\ldots$.  

\section{The principal symbol and its symmetries}

\label{sec:princ}

\subsection{Definition}

Consider the theory defined by the action \eqref{action1}. For the moment we assume only that the metric $g_{\mu\nu}$ is non-degenerate but we do not require it to have a particular signature. In particular, the following analysis applies for both Lorentzian and Riemannian signature. Define
\be
 E^{\mu\nu} = - \frac{16\pi G}{\sqrt{|g|}} \frac{\delta S}{\delta g_{\mu\nu}} \qquad \qquad E^I =  - \frac{16\pi G}{\sqrt{|g|}} \frac{\delta S}{\delta \phi_I}
\ee
so the equations of motion are
\be
 E^{\mu\nu} = E^I = 0
\ee
We assume that these equations of motion are second order. 

Let $\xi_\mu$ be an arbitrary covector. Then the principal symbol is defined as in \eqref{PS_def} where 
\ba
P_{gg}^{\mu\nu\rho\sigma\alpha\beta}  &\equiv& \frac{\partial E^{\mu\nu}}{\partial (\partial_\alpha\partial_\beta g_{\rho \sigma})} \qquad \qquad
P_{gm}^{\mu\nu I \alpha \beta} \equiv \frac{\partial E^{\mu\nu}}{\partial (\partial_\alpha\partial_\beta \phi_I)}
\nonumber \\
P_{mg}^{I \mu \nu\alpha\beta} &\equiv& \frac{\partial E^I}{\partial (\partial_\alpha\partial_\beta g_{\mu\nu})} 
\qquad \qquad
P_{mm}^{IJ\alpha\beta}  \equiv \frac{\partial E^I}{\partial (\partial_\alpha\partial_\beta \phi_J)} 
\ea
These definitions are equivalent to those obtained by linearising the equations of motion as in \eqref{lin_eqs}. By definition these objects possess the symmetries
\be
\label{Psym0}
P_{gg}^{\mu\nu\rho\sigma\alpha\beta}=P_{gg}^{(\mu\nu)\rho\sigma\alpha\beta}=P_{gg}^{\mu\nu(\rho\sigma)\alpha\beta}=P_{gg}^{\mu\nu\rho\sigma(\alpha\beta)}
\ee
\be
\label{Pgmsym0}
P_{gm}^{\mu\nu I \alpha \beta} = P_{gm}^{(\mu\nu) I \alpha \beta} = P_{gm}^{\mu\nu I (\alpha \beta)}
\ee
\be
 P_{mg}^{I \mu \nu\alpha\beta} = P_{mg}^{I (\mu \nu)\alpha\beta}=P_{mg}^{I \mu \nu(\alpha\beta)}
\ee
\be
 P_{mm}^{IJ\alpha\beta} = P_{mm}^{IJ(\alpha\beta)}
\ee
We will sometimes use the notation
\ba
 P_{gg}^{\mu\nu\rho\sigma}(\xi) &\equiv&  P_{gg}^{\mu\nu\rho\sigma\alpha\beta}\xi_\alpha \xi_\beta \qquad \qquad
 P_{gm}^{\mu\nu I}(\xi) \equiv P_{gm}^{\mu\nu I \alpha \beta}\xi_\alpha \xi_\beta
\nonumber \\ 
P_{mg}^{I \mu \nu}(\xi) &\equiv&  P_{mg}^{I \mu \nu\alpha\beta}\xi_\alpha \xi_\beta \qquad \qquad
P_{mm}^{IJ}(\xi) \equiv P_{mm}^{IJ\alpha\beta} \xi_\alpha \xi_\beta
\ea
The principal symbol ${\cal P}(\xi)$ is a matrix that acts on the vector space of ``polarization" vectors of the form $(t_{\mu\nu},t_I)$ where $t_{\mu\nu}$ is symmetric. In a general theory, ${\cal P}(\xi)$ depends on the fields $(g_{\mu\nu},\phi_I)$ and their first and second derivatives.

\subsection{Consequences of the action principle}

We will now show that the action principle implies that the principal symbol is a symmetric matrix. Fix a ``background" field configuration (not necessarily a solution) and consider a 2-parameter compactly supported variation of this configuration, parameterized by $(\lambda_1,\lambda_2)$. This means that we consider fields $g_{\mu\nu}(x,\lambda_1,\lambda_2)$ and $\phi_I(x,\lambda_1,\lambda_2)$ which coincide with the background fields $g_{\mu\nu}(x,0,0)$ and $\phi_I(x,0,0)$ outside a compact set. Hence $\partial_{\lambda_i} g_{\mu\nu}$ and $\partial_{\lambda_i} \phi_I$ are functions of compact support in spacetime. 

Take a derivative w.r.t. $\lambda_1$ of the action evaluated on these fields:
\be
\label{deltaS}
\delta_1 S= -\frac{1}{16\pi G} \int d^d x \sqrt{|g|} \left( E^{\mu\nu}\delta_1 g_{\mu\nu} + E^I \delta_1 \phi_I \right)
\ee
where $\delta_1$ denotes a partial derivative w.r.t. $\lambda_1$ and compact support lets us discard total derivatives. Now take a derivative w.r.t. $\lambda_2$ to obtain
\ba
 \delta_2 \delta_1 S &=& -\frac{1}{16\pi G}  \int d^d x \sqrt{|g|} \left[E^{\mu\nu}\delta_2 \delta_1 g_{\mu\nu} + E^I\delta_2 \delta_1 \phi_I 
\right.\nonumber \\ 
 &+& \left. \left( P_{gg}^{\mu\nu\rho\sigma\alpha\beta}\partial_\alpha \partial_\beta \delta_2 g_{\rho\sigma} + P_{gm}^{\mu\nu I\alpha\beta} \partial_\alpha\partial_\beta \delta_2 \phi_I +\ldots \right) \delta_1 g_{\mu\nu} 
\right.\nonumber \\
 &+&\left.  \left( P_{mg}^{I \mu\nu \alpha \beta}\partial_\alpha \partial_\beta \delta_2 g_{\mu\nu} + P_{mm}^{IJ \alpha \beta}\partial_\alpha \partial_\beta \delta_2 \phi_J +\ldots \right) \delta_1 \phi_I 
\right]
 \ea
where the ellipses denotes terms with fewer than $2$ derivatives acting on $\delta_2 g_{\mu\nu}$ and $\delta_2 \phi_I$. Since we are not keeping track of such terms, we can replace partial derivatives above with covariant derivatives. Integrating by parts we have
\ba
\label{delta1delta2}
\delta_2 \delta_1 S &=& \frac{1}{16\pi G}  \int d^d x \sqrt{|g|} \left( -E^{\mu\nu}\delta_2 \delta_1 g_{\mu\nu} - E^I\delta_2 \delta_1 \phi_I 
+  P_{gg}^{\mu\nu\rho\sigma\alpha\beta}  \nabla_\alpha \delta_1 g_{\mu\nu} \nabla_\beta \delta_2 g_{\rho\sigma} 
\right.\nonumber \\ 
&+&  P_{gm}^{\mu\nu I\alpha\beta} \nabla_\alpha \delta_1 g_{\mu\nu} \nabla_\beta \delta_2 \phi_I
+ P_{mg}^{I \mu\nu \alpha \beta}\nabla_\alpha \delta_1 \phi_I \nabla_\beta \delta_2 g_{\mu\nu} + P_{mm}^{IJ \alpha \beta}\nabla_\alpha \delta_1 \phi_I \nabla_\beta \delta_2 \phi_J 
\nonumber \\
&+& \left. \ldots \right)
\ea
where the ellipsis denotes terms for which the total number of spacetime derivatives acting on the field variations is less than $2$. 
Now antisymmetrize in $\lambda_1$ and $\lambda_2$ and evaluate at $\lambda_1=\lambda_2=0$: the LHS vanishes and terms involving $\delta_2 \delta_1 g_{\mu\nu}$ and $\delta_2 \delta_1 \phi_I$ drop out of the RHS when antisymmetrized, leaving (after using the symmetry of the principal symbol on $\alpha\beta$)
\ba
 0 = \frac{1}{16\pi G} \int d^d x &\sqrt{|g|}& \left\{ \left( P_{gg}^{\mu\nu\rho\sigma\alpha\beta} - P_{gg}^{\rho\sigma \mu\nu\alpha \beta}\right) \nabla_\alpha \delta_1 g_{\mu\nu} \nabla_\beta \delta_2 g_{\rho\sigma} \right. \nonumber \\
&+& \left[  \left( P_{gm}^{\mu\nu I\alpha\beta} - P_{mg}^{I \mu\nu\alpha \beta}\right)\nabla_\alpha \delta_1 g_{\mu\nu}\nabla_\beta \delta_2 \phi_I + (1 \leftrightarrow 2)  \right] \nonumber \\
&+& \left. \left( P_{mm}^{IJ\alpha \beta} - P_{mm}^{JI \alpha \beta} \right) \nabla_\alpha \delta_1 \phi_I \nabla_\beta \delta_2 \phi_J + \ldots \right\} 
\ea
This has to hold for arbitrary compactly supported variations. Hence the coefficients of the terms quadratic in first derivatives of the variations must vanish:
\be
\label{Psym1}
 P_{gg}^{\mu\nu\rho\sigma\alpha\beta} = P_{gg}^{\rho\sigma \mu\nu\alpha \beta} \qquad \qquad P_{gm}^{\mu\nu I\alpha\beta} = P_{mg}^{I \mu\nu\alpha \beta}\qquad \qquad  P_{mm}^{IJ\alpha \beta} = P_{mm}^{JI \alpha \beta}
\ee
These equations are equivalent to the statement that the principal symbol \eqref{PS_def} is symmetric. The equations of motion were not used in the above argument, so this holds for any field configuration. 

If the background {\it does} satisfy the equations of motion then by setting $\delta_2 = \delta_1$ in \eqref{delta1delta2} we see that the second variation of the action around a solution is
\ba
\label{2ndvariation}
 \delta_1^2  S &=& \frac{1}{16\pi G}  \int d^d x \sqrt{|g|} \left(  P_{gg}^{\mu\nu\rho\sigma\alpha\beta}  \nabla_\alpha \delta_1 g_{\mu\nu} \nabla_\beta \delta_1 g_{\rho\sigma} 
+  P_{gm}^{\mu\nu I\alpha\beta} \nabla_\alpha \delta_1 g_{\mu\nu} \nabla_\beta \delta_1 \phi_I
\right.\nonumber \\ 
&+& \left. P_{mg}^{I \mu\nu \alpha \beta}\nabla_\alpha \delta_1 \phi_I \nabla_\beta \delta_1 g_{\mu\nu} + P_{mm}^{IJ \alpha \beta}\nabla_\alpha \delta_1 \phi_I \nabla_\beta \delta_1 \phi_J
+  \ldots \right)
\ea
where the ellipsis denotes terms for which the total number of derivatives acting on field variations is less than $2$. Thus the principal symbol determines the 2-derivative terms in the action when expanded to quadratic order around a background solution. 

\subsection{Consequences of diffeomorphism invariance}

Next we will establish some further symmetries that follow from invariance of the action under compactly supported diffeomorphisms. Consider an infinitesimal diffeomorphism generated by a compactly supported vector field $X^\mu$. We assume that the matter fields are such that their transformation does not involve derivatives of $X^\mu$:
\be
\label{matter_diff}
 \delta g_{\mu\nu} = 2 \nabla_{(\mu} X_{\nu)} \qquad \qquad \delta \phi_I = X^\mu L_{\mu I} 
\ee
for some tensor fields $L_{\mu I}$. For example, a set of scalar fields $\phi_I$ has $L_{\mu I} = \partial_\mu \phi_I$ and a vector potential  $A_\mu$ has $\delta A_\nu = X^\mu F_{\mu\nu}$ (up to a compactly supported gauge transformation) so $L_{\mu\nu} = F_{\mu\nu}$. Taking the variation in \eqref{deltaS} to be such a diffeomorphism, with compactly supported $X^\mu$, the LHS vanishes and integrating by parts leads to the generalized Bianchi identity:
\be
 \nabla_\nu E^{\mu\nu} - L^\mu{}_I E^I = 0
\ee 
This has to hold for an arbitrary field configuration. Using the chain rule to expand the first term gives
\be
 P_{gg}^{\mu\nu\rho\sigma\alpha\beta} \partial_\nu \partial_\alpha \partial_\beta g_{\rho\sigma} + P_{gm}^{\mu\nu I \alpha \beta} \partial_\nu \partial_\alpha \partial_\beta\phi_I  + \ldots =0 
\ee
where the ellipsis denotes terms not involving third (or higher) derivatives of the fields. Since this equation has to hold for an arbitrary field configuration, the coefficients of the third derivative terms must vanish, which requires
\be
 P_{gg}^{\mu(\nu|\rho\sigma|\alpha \beta)}=0 \qquad \qquad P_{gm}^{\mu(\nu| I |\alpha \beta)}=0
\ee
Using the symmetries \eqref{Psym1} these equations are equivalent to
\be
\label{Psym2}
P_{gg}^{\mu\nu\rho(\sigma\alpha \beta)}=0 \qquad \qquad P_{mg}^{I\mu(\nu\alpha \beta)}=0
\ee

\subsection{Combining the symmetries}

\label{sec:princ_sym}

We will now use the results established above to prove the following proposition.

\medskip

\noindent {\bf Proposition.} 
{\it 
Assume that the action \eqref{action1} is invariant under compactly supported infinitesimal diffeomorphisms which act on the matter fields as in \eqref{matter_diff}. 
Then the components of the principal symbol can be written (for arbitrary $\xi_\mu$)
\be
\label{PCI}
 P_{mg}^{I\mu\nu}(\xi) = C^{I \mu\alpha\nu\beta}\xi_\alpha \xi_\beta
\ee
\be
\label{PggC}
 P_{gg}^{\mu\nu\rho\sigma}(\xi) = C^{\mu (\rho|\alpha\, \nu|\sigma) \beta}\xi_\alpha \xi_\beta
\ee
where $C^{I\mu\nu\rho\sigma}$ has the symmetries of a metric-derived Riemann tensor:
\be
\label{CIsymall}
 C^{I\mu\nu\rho\sigma} = C^{I[\mu\nu]\rho\sigma} = C^{I\mu\nu[\rho\sigma]} = C^{I\rho\sigma\mu\nu} \qquad C^{I\mu[\nu\rho\sigma]}=0
\ee
and $C^{\mu_1\mu_2\mu_3 \, \nu_1 \nu_2\nu_3}$ has the symmetries
\be
\label{Csymall}
C^{\mu_1\mu_2\mu_3 \, \nu_1 \nu_2\nu_3}=C^{[\mu_1\mu_2\mu_3] \, \nu_1 \nu_2\nu_3}=C^{\mu_1\mu_2\mu_3 \, [\nu_1 \nu_2\nu_3]}=C^{\nu_1 \nu_2\nu_3\, \mu_1\mu_2\mu_3}
\ee 
and
\be
\label{Csymbianchi}
 C^{\mu_1\mu_2[\mu_3 \, \nu_1 \nu_2\nu_3]} = C^{\mu_1[ \mu_2 \mu_3 \, \nu_1 \nu_2]\nu_3}=0
\ee
The tensors $C^{I\mu\nu\rho\sigma}$ and $C^{\mu_1\mu_2\mu_3 \, \nu_1 \nu_2\nu_3}$ are uniquely defined by the above properties. They depend on the fields $(g_{\mu\nu},\phi_I)$ and their first and second derivatives. 
}

\medskip

\noindent {\it Proof.} First note that
\ba
 P_{mg}^{I \mu \nu \alpha \beta} &=& -P_{mg}^{I \mu \alpha \beta \nu} - P_{mg}^{I \mu \beta\nu \alpha} = -P_{mg}^{I \alpha \mu \beta\nu} - P_{mg}^{I \beta\mu\nu\alpha} \nonumber \\
 &=& P_{mg}^{I \alpha \beta\nu\mu}+P_{mg}^{I\alpha \nu \mu\beta}+P_{mg}^{I \beta\nu\alpha\mu}+P_{mg}^{I \beta\alpha \mu\nu} \nonumber \\
 &=& 2 P_{mg}^{I \alpha \beta\mu\nu}+P_{mg}^{I\nu\alpha\mu\beta}+P_{mg}^{I\nu\beta\alpha\mu}=2 P_{mg}^{I \alpha \beta\mu\nu}-P_{mg}^{I\nu\mu\alpha\beta}
\ea
where we have used \eqref{Psym2} and \eqref{Pgmsym0} repeatedly. Rearranging we have
\be
\label{Pmgsym1}
 P_{mg}^{I\mu\nu\alpha\beta} = P_{mg}^{I \alpha \beta \mu \nu}
\ee
i.e. $P_{mg}^{I\mu\nu\alpha\beta}$ is symmetric under interchange of the pair $\mu\nu$ with the pair $\alpha \beta$. It now follows from \eqref{Psym2} that
\be
 P_{mg}^{I(\mu\nu\alpha)\beta}=P_{mg}^{I\beta(\alpha \mu\nu)}=0
\ee
so $P_{mg}^{I\mu\nu\alpha\beta}$ vanishes when symmetrized on any three Greek indices. We now define
\be
\label{CIdef}
 C^{I\mu\alpha\nu\beta}=\frac{2}{3}\left( P_{mg}^{I\mu\nu\alpha\beta}-P_{mg}^{I\mu\beta\alpha\nu} \right)
\ee
The symmetries \eqref{Pmgsym1} and \eqref{Pgmsym0} imply that
\be
\label{CIsym}
 C^{I \mu\alpha\nu\beta} = C^{I \nu \beta \mu \alpha}
\ee
Furthermore we have
\be
 C^{I\mu\alpha\nu\beta}=C^{I\mu\alpha[\nu\beta]} =C^{I[\mu\alpha]\nu\beta}
\ee
where the first equality follows from the definition of the LHS and the second equality then follows from \eqref{CIsym}. Finally, the symmetry \eqref{Pgmsym0} implies
\be
 C^{I\mu[\alpha \nu \beta]}=0
\ee
Thus $C^{I\mu\alpha\nu\beta}$ has the Riemann symmetries stated in \eqref{CIsym}. Now consider
\be
 C^{I \mu (\alpha |\nu|\beta)}=\frac{2}{3} P_{mg}^{I\mu\nu\alpha\beta} - \frac{1}{3} P_{mg}^{I \mu\beta\alpha \nu}-\frac{1}{3} P_{mg}^{I \mu\alpha\nu \beta} = P_{mg}^{I\mu\nu\alpha\beta}
\ee
using \eqref{Psym2} in the final equality. Hence the relation between $C^I$ and $P_{mg}^I$ is invertible. In particular we have established equation \eqref{PCI}. Conversely, to establish uniqueness of $C^I$ with the above symmetries and satisfying \eqref{PCI}, note that the symmetries imply that $C^{I\mu\nu\rho\sigma} = (2/3)(C^{I \mu (\nu|\rho|\sigma)} - C^{I \mu (\nu|\sigma|\rho)})$ and using \eqref{PCI} this reduces to \eqref{CIdef}. 

We can apply very similar arguments to $P_{gg}$. First we have
\ba
 P_{gg}^{\mu\nu\rho\sigma\alpha \beta}&=& - P_{gg}^{\mu\nu\rho\alpha \beta\sigma} - P_{gg}^{\mu\nu\rho\beta \alpha\sigma} =  - P_{gg}^{\mu\nu\alpha\rho \beta\sigma} - P_{gg}^{\mu\nu\beta\rho \alpha\sigma} \nonumber \\
&=&  P_{gg}^{\mu\nu\alpha\beta \rho\sigma} + P_{gg}^{\mu\nu\alpha\sigma \beta \rho} + P_{gg}^{\mu\nu\beta\alpha \rho\sigma} + P_{gg}^{\mu\nu\beta\sigma \rho \alpha} \nonumber \\
&=& 2 P_{gg}^{\mu\nu \alpha \beta \rho \sigma} + P_{gg}^{\mu\nu\sigma \alpha \beta \rho}+ P_{gg}^{\mu\nu\sigma\beta \rho \alpha} = 2 P_{gg}^{\mu\nu \alpha \beta \rho \sigma} - P_{gg}^{\mu\nu\sigma\rho\alpha \beta}
\ea
where we have used \eqref{Psym2} and \eqref{Psym0} repeatedly. Rearranging we have
\be
  P_{gg}^{\mu\nu\rho\sigma\alpha \beta} = P_{gg}^{\mu\nu \alpha \beta \rho \sigma}
\ee
and so combining with \eqref{Psym1} we see that $P_{gg}^{\mu\nu\rho\sigma\alpha\beta}$ has a ``pairwise interchange" symmetry, i.e., it is symmetric under interchange of any of the pairs $\mu\nu$, $\rho\sigma$ and $\alpha \beta$. 

Now consider
\ba
 6P_{gg}^{\mu(\nu \rho |\sigma |\alpha) \beta} &=& \left( P_{gg} ^{\mu\nu\alpha\sigma\rho \beta} + P_{gg} ^{\mu\nu\rho\sigma\alpha \beta}  \right) +  \left( P_{gg} ^{\mu\rho\nu\sigma\alpha \beta} + P_{gg} ^{\mu\rho\alpha \sigma\nu \beta}  \right) +  \left( P_{gg} ^{\mu\alpha\rho\sigma\nu \beta} + P_{gg} ^{\mu\alpha\nu\sigma\rho \beta}  \right) \nonumber \\
 &=& - P_{gg}^{\mu\nu\sigma\beta\alpha\rho} - P_{gg}^{\mu\rho\sigma\beta\alpha\nu}-P_{gg}^{\mu\alpha \sigma \beta \nu\rho}\nonumber \\ &=&- P_{gg}^{\sigma\beta\mu\nu\alpha\rho} - P_{gg}^{\sigma\beta\mu\rho\alpha\nu}-P_{gg}^{ \sigma \beta\mu\alpha \nu\rho}=0
\ea
where the second equality uses \eqref{Psym0} and \eqref{Psym2}, the third equality uses \eqref{Psym1} and the final equality is \eqref{Psym2}. From the results that we have obtained, it follows that $P_{gg}^{\mu\nu\rho\sigma\alpha\beta}$ vanishes when symmetrized on any three indices.\footnote{This implies that $P_{gg}$ is identically zero for $d \le 2$ dimensions (since at least $3$ indices must take the same value for $d \le 2$).}

We now define
\be
\label{CdefPgg}
C^{\mu \rho \alpha \, \nu \sigma \beta} = \frac{2}{3} \left( P_{gg}^{\mu\nu\rho\sigma\alpha \beta} + P_{gg}^{\rho\nu \alpha \sigma\mu \beta} + P_{gg}^{\alpha\nu\mu\sigma\rho\beta} \right)
\ee
The symmetries \eqref{Psym0} imply that $C^{\mu \rho \alpha \, \nu \sigma \beta}$ vanishes when antisymmetrized on any four indices, so \eqref{Csymbianchi} holds. The pairwise interchange symmetry of $P_{gg}$ implies that
\be
 C^{\mu \rho \alpha\,  \nu \sigma \beta} = C^{ \nu \sigma \beta\,  \mu \rho \alpha}
\ee
so $C$ is symmetric under interchange of the first three indices with the final three indices.
We can write the definition of $C$ as
\be
 C^{\mu \rho \alpha\,  \nu \sigma \beta}=2\left( P_{gg}^{[\mu|\nu|\rho|\sigma|\alpha]\beta} + P_{gg}^{(\mu|\nu|\rho|\sigma|\alpha)\beta}  \right)
\ee
i.e. the first term is antisymmetrized on indices $\mu\rho\alpha$ and the second term is symmetrized on indices $\mu\rho\alpha$. But we've seen that the symmetrized term vanishes. It follows that $C$ is totally antisymmetric on its first three indices and also on its final three indices:
\be
 C^{\mu \rho \alpha\,  \nu \sigma \beta} = C^{[\mu \rho \alpha] \,  \nu \sigma \beta}=C^{\mu \rho \alpha\, [ \nu \sigma \beta]}
\ee
where the second equality follows from the interchange symmetry. We have now established the symmetries \eqref{Csymall}. 

Next consider the symmetrization of $C^{\mu \rho \alpha\,  \nu \sigma \beta}$ on $\alpha \beta$ and on $\rho\sigma$. Using the symmetries of $P_{gg}$ we obtain
\ba
\label{PggC2}
 &&\frac{1}{4}\left( C^{\mu \rho \alpha\,  \nu \sigma \beta} + C^{\mu \sigma\beta\,  \nu \rho \alpha} + C^{\mu \sigma \alpha\,  \nu \rho \beta} + C^{\mu \rho \beta\,  \nu \sigma \alpha} \right)  \nonumber \\
 &=&
 \scriptstyle{
 \frac{1}{6} \left( 4P_{gg}^{\mu\nu\rho\sigma\alpha \beta} + 
 P_{gg}^{\rho\nu \alpha \sigma\mu \beta}+P_{gg}^{\rho\nu \beta \sigma\mu \alpha}+P_{gg}^{\sigma\nu \alpha \rho\mu \beta}+P_{gg}^{\sigma\nu \beta \rho\mu \alpha}+ P_{gg}^{\alpha\nu\mu\sigma\rho\beta}+ P_{gg}^{\beta\nu\mu\sigma\rho\alpha}+  P_{gg}^{\alpha\nu\mu\rho\sigma\beta}+ P_{gg}^{\beta\nu\mu\rho\sigma\alpha}\right)
}
  \nonumber \\
 &=& \frac{1}{6} \left( 4P_{gg}^{\mu\nu\rho\sigma\alpha \beta} -P_{gg}^{\rho\nu\mu\sigma\alpha \beta} -P_{gg}^{\sigma\nu\mu\rho\alpha\beta} -P_{gg}^{\rho\nu \mu\sigma \alpha\beta} - P_{gg}^{\sigma\nu\mu\rho \alpha \beta}\right)\nonumber \\
 &=& \frac{1}{6} \left( 4P_{gg}^{\mu\nu\rho\sigma\alpha \beta} -2P_{gg}^{\rho\nu\mu\sigma\alpha \beta} -2P_{gg}^{\sigma\nu\mu\rho\alpha\beta} \right) \nonumber \\
 &=& \frac{1}{6} \left( 4P_{gg}^{\mu\nu\rho\sigma\alpha \beta} +2P_{gg}^{\rho\sigma\mu\nu\alpha \beta}  \right) = P_{gg}^{\mu\nu\rho\sigma\alpha \beta}
\ea
The first equality is the definition of $C$. We then apply the 3-index symmetrization property of $P_{gg}$ to the second and third terms, the fourth and fifth terms etc to obtain the second equality. We then apply this symmetrization property to the second and fourth terms and to the third and fifth terms to obtain the third equality. Using the symmetrization property again gives the fourth equality, and the final equality then follows from the pairwise interchange symmetry. This result shows that we can invert the relation between $C$ and $P_{gg}$ so these two tensors contain the same information. Equation \eqref{PggC2} is equivalent to \eqref{PggC} and so we have demonstrated that \eqref{CdefPgg} satisfies all of the properties listed in the proposition.

Finally we will demonstrate that \eqref{CdefPgg} is the unique expression for $C^{\mu \rho \alpha \, \nu \sigma \beta}$ that satisfies the properties listed in the proposition. By linearity, this is equivalent to showing that if $C^{\mu \rho \alpha \, \nu \sigma \beta}$ has the symmetries \eqref{Csymall} and \eqref{Csymbianchi} and the RHS of \eqref{PggC} vanishes (for any $\xi_\mu$) then $C^{\mu \rho \alpha \, \nu \sigma \beta}$ must vanish. So assume \eqref{Csymall} and \eqref{Csymbianchi} and that the RHS of \eqref{PggC} vanishes. The latter equation gives
\be
 C^{\mu\rho \alpha \, \nu \sigma \beta} + C^{\mu \sigma \beta \, \nu \rho \alpha} + C^{\mu\sigma \alpha \, \nu \rho \beta}+C^{\mu\rho \beta \, \nu \sigma \alpha}=0
\ee
Now add to this equation the two equations obtained by cycling the indices $\mu\rho\alpha$ (this is motivated by the RHS of \eqref{CdefPgg}). The result is
\ba
 0&=&
 \scriptstyle{
 3 C^{\mu\rho \alpha \, \nu \sigma \beta} + C^{\mu \sigma \beta \, \nu \rho \alpha} + C^{\mu\sigma \alpha \, \nu \rho \beta}+C^{\mu\rho \beta \, \nu \sigma \alpha}+C^{\rho\sigma \beta\, \nu\alpha \mu}+C^{\rho\sigma\mu\, \nu\alpha\beta}+C^{\rho\alpha\beta\, \nu \sigma \mu}+C^{\alpha\sigma\beta\, \nu\mu\rho}+C^{\alpha\sigma\rho\,\nu\mu\beta} + C^{\alpha\mu\beta\, \nu\sigma \rho}
} \nonumber \\
&=&
\scriptstyle{
3 C^{\mu\rho \alpha \, \nu \sigma \beta} + C^{\mu \sigma \beta \, \nu \rho \alpha} + C^{\mu\sigma \alpha \, \nu \rho \beta}+C^{\mu\rho \beta \, \nu \sigma \alpha}+C^{\mu \nu\alpha \,\rho\sigma \beta}+C^{\mu\rho\sigma\, \nu\alpha\beta}+
C^{  \mu\nu \sigma\, \rho\alpha\beta}+C^{\mu\rho \nu\, \alpha\sigma\beta}+C^{\mu\beta \nu\, \alpha\sigma\rho} + C^{\mu\beta \alpha\, \nu\sigma \rho}
}\nonumber \\
&=&
4 C^{\mu\rho \alpha \, \nu \sigma \beta}+C^{\mu \sigma \beta \, \nu \rho \alpha}+C^{\mu\rho \beta \, \nu \sigma \alpha}+2 C^{  \mu\nu \sigma\, \rho\alpha\beta}+C^{\mu\beta \nu\, \alpha\sigma\rho} + C^{\mu\beta \alpha\, \nu\sigma \rho}
\nonumber \\
&=& 4 C^{\mu\rho \alpha \, \nu \sigma \beta}-C^{\mu \beta  \sigma \, \nu \rho \alpha}-C^{\mu \beta \rho\, \nu \sigma \alpha}+2 C^{  \mu\nu \sigma\, \rho\alpha\beta}+C^{\mu\beta \nu\, \alpha\sigma\rho} + C^{\mu\beta \alpha\, \nu\sigma \rho}
\nonumber \\
&=& 4 C^{\mu\rho \alpha \, \nu \sigma \beta} - 2C^{\mu \beta  \sigma \, \nu \rho \alpha}+2 C^{  \mu\nu \sigma\, \rho\alpha\beta}+2C^{\mu\beta \nu\, \alpha\sigma\rho} 
\nonumber \\
&=& 4 C^{\mu\rho \alpha \, \nu \sigma \beta}+2C^{\mu \beta  \sigma \, \nu  \alpha \rho}-2 C^{  \mu\nu \sigma\, \beta\alpha\rho}-2C^{\mu\beta \nu\, \sigma\alpha\rho}
\nonumber \\
&=& 4 C^{\mu\rho \alpha \, \nu \sigma \beta}+2C^{\nu\beta\sigma\, \mu \alpha \rho}
\nonumber \\
&=& 6C^{\mu\rho \alpha \, \nu \sigma \beta}
\ea 
In the first and second equalities we have used the symmetries \eqref{Csymall}. In the third equality we used the symmetry $C^{\mu[\rho\alpha\nu\sigma]\beta}=0$ from \eqref{Csymbianchi}. The fourth equality uses  \eqref{Csymall}. In the fifth equality we used $C^{\mu\beta[\sigma \nu\rho\alpha]}=0$ from \eqref{Csymbianchi}. The sixth equality uses  \eqref{Csymall}. The seventh equality uses $C^{[\mu \beta \sigma \nu] \alpha \rho}=0$ from combining \eqref{Csymall} with \eqref{Csymbianchi}. The final equality uses \eqref{Csymall}. This concludes the proof.

\subsection{Example: Einstein-Hilbert action}

As an example, consider, for Lorentzian signature in any number of dimensions, the Einstein-Hilbert action $L=R$ and no matter fields. We have $E^{\mu\nu} = R^{\mu\nu} - (1/2) R g^{\mu\nu}$. The principal symbol (acting on $t_{\mu\nu}$) can be read off by the substitution $\partial_\mu\partial_\nu g_{\rho \sigma} \rightarrow \xi_\mu \xi_\nu t_{\rho\sigma}$ with the result
\be
\label{EH1}
 (P_{gg})_{\mu\nu}{}^{\rho\sigma}(\xi) t_{\rho\sigma} = \left( -\frac{1}{2} \xi^\rho \xi_\rho \tilde{t}_{\mu\nu} + \xi^\rho \xi_{(\mu}\tilde{t}_{\nu) \rho}-\frac{1}{2} g_{\mu\nu}\xi^\rho \xi^\sigma \tilde{t}_{\rho\sigma}\right) \qquad \tilde{t}_{\mu\nu} \equiv t_{\mu\nu} - \frac{1}{2} g_{\mu\nu} t^\rho_\rho
\ee
This can be rewritten as
\be
 (P_{gg})_{\mu}{}^{\nu\rho\sigma}(\xi) t_{\rho\sigma} = \frac{1}{2}\delta_{\mu\rho \alpha}^{\nu\sigma \beta} t^\rho_\sigma \xi^\alpha \xi_\beta
\ee
where
\be
\label{gendeltadef}
 \delta^{\mu_1 \ldots \mu_q}_{\nu_1 \ldots \nu_q} \equiv q! \delta^{\mu_1}_{[\nu_1} \ldots \delta^{\mu_q}_{\nu_q]} 
\ee
hence for the Einstein-Hilbert action we have
\be
\label{Cdelta}
 C_{\mu\rho\alpha}{}^{\nu\sigma\beta}=  \frac{1}{2}\delta_{\mu\rho \alpha}^{\nu\sigma \beta} 
\ee
which clearly has the symmetries we have described above.

The result \eqref{Cdelta} also holds if we include ``minimally coupled" matter. More precisely it holds for $L =  R + L_m$ if $L_m$ does not produce terms in $E^{\mu\nu}$ involving second derivatives of the metric (so $L_m$ does not affect $P_{gg}$).  

\subsection{Example: Lovelock theories}

The principal symbol for Lovelock theories was calculated in \cite{Aragone:1987jm,Choquet-Bruhat1988,Reall2014,Reall2014a}. From the results of \cite{Reall2014a} we can read off an expression that satisfies \eqref{PggC}:
\be
\label{lovelock}
 C_{\mu\rho\alpha}{}^{\nu\sigma\beta}= -2\sum\limits_{p\geq 1} p~ k_p~ \delta_{\mu\rho\alpha\beta_1\beta_2...\beta_{2p-3}\beta_{2p-2}}^{\nu\sigma\beta\alpha_1\alpha_2...\alpha_{2p-3}\alpha_{2p-2}} R_{\alpha_1\alpha_2}{}^{\beta_1\beta_2}...~R_{\alpha_{2p-3}\alpha_{2p-2}}{}^{\beta_{2p-3}\beta_{2p-2}}.
\ee 
where the $k_p$ are the coupling constants of the theory and the antisymmetry ensures that the sum is finite. With a little work one can check that this expression satisfies \eqref{Csymall} and \eqref{Csymbianchi}. Hence, by uniqueness, this must be the correct result. Taking $k_1=-1/4$ gives a conventionally normalised Einstein term, i.e., it ensures that the $p=1$ term agrees with \eqref{Cdelta}. 

\subsection{Maxwell field}

\label{sec:maxwell}

Let's consider the case in which our ``matter" field is a Maxwell vector potential, i.e., $\phi_I \rightarrow A_\mu$, so we replace indices $I,J,\ldots$ with $\mu,\nu,\ldots$. We assume that the action is invariant under compactly supported gauge transformations.\footnote{
Note that Chern-Simons terms such as $\int A \wedge F$ (for $d=3$) or $\int A \wedge F \wedge F$ (for $d=5$) are invariant under gauge transformations of compact support so our analysis applies to theories containing such terms.} We can use gauge invariance to deduce some further symmetries of the principal symbol. Take $\delta_1$ to be a compactly supported gauge transformation in \eqref{deltaS}. Gauge invariance of the action, and $\delta_1 g_{\mu\nu}=0$ then imply the Bianchi identity
\be
 0=\nabla_\mu E^\mu=P_{mg}^{\mu \rho\sigma \alpha \beta} \partial_\mu \partial_\alpha \partial_\beta g_{\rho\sigma} + P_{mm}^{\mu\nu\alpha\beta}\partial_\mu \partial_\alpha \partial_\beta A_\nu + \ldots
\ee
where the ellipsis denotes terms not involving third (or higher) derivatives. This identity has to hold for any field configuration, so the third derivative terms must vanish, which requires
\be
 P_{mg}^{(\mu|\rho\sigma|\alpha \beta)}=0 
\ee
and (using symmetry on $\mu\nu$)
\be
 P_{mm}^{\mu(\nu \alpha \beta)}=0
\ee
We also note the second equation of \eqref{Psym2}, which is
\be
 P_{mg}^{\mu\rho(\sigma\alpha \beta)}=0 
\ee
Similar arguments to those used in section \ref{sec:princ_sym} imply that we can write
\be
 P_{mm}^{\mu\nu\alpha\beta}=M^{\mu (\alpha|\nu|\beta)}
\ee
where $M^{\mu\alpha\nu\beta}$ has the symmetries of a (metric induced) Riemann tensor. 

\subsection{Four dimensions}

\label{sec:4d}

Now return to a general theory with $d=4$ and assume that $g_{\mu\nu}$ has Lorentzian signature. The symmetries of $C^{\mu_1 \mu_2 \mu_3 \, \nu_1 \nu_2 \nu_3}$ imply that we can define a symmetric tensor $C_{\mu\nu}$ by 
\be
\label{Cdef}
 C^{\mu_1 \mu_2 \mu_3 \, \nu_1 \nu_2 \nu_3} =-\frac{1}{2} \epsilon^{\mu_1 \mu_2 \mu_3 \rho}\epsilon^{\nu_1 \nu_2 \nu_3 \sigma} C_{\rho\sigma}  
\ee
where $\epsilon_{\mu\nu\rho\sigma}$ is the volume form on spacetime defined by the metric. In a (right handed) basis we have 
\be
\epsilon^{\mu\nu\rho\sigma} = \frac{1}{\sqrt{-g}} \tilde{\epsilon} ^{\mu\nu\rho\sigma}
\ee
where $\tilde{\epsilon}^{\mu\nu\rho\sigma}$ is a totally antisymmetric tensor density with $\tilde{\epsilon}^{0123}=- 1$. 

For example, consider the Einstein-Hilbert Lagrangian $L = R$. Recall that
\be
\label{delta_eps}
 \delta^{\mu_1 \mu_2 \mu_3}_{\nu_1\nu_2\nu_3} = - \epsilon^{\mu_1\mu_2\mu_3 \rho}\epsilon_{\nu_1\nu_2 \nu_3 \rho}
\ee
hence equation \eqref{Cdelta} can be rewritten
\be
\label{Cdelta2}
C^{\mu_1 \mu_2 \mu_3 \, \nu_1 \nu_2 \nu_3}= - \frac{1}{2}\epsilon^{\mu_1 \mu_2 \mu_3 \rho}\epsilon^{\nu_1 \nu_2 \nu_3 \sigma} g_{\rho\sigma} \qquad \qquad {\rm (Einstein-Hilbert)}
\ee
so we have $C_{\mu\nu} = g_{\mu\nu}$ for the Einstein-Hilbert action. This result also holds if we include minimally coupled matter fields.

Henceforth we will restrict attention to the case for which $C_{\mu\nu}$ is non-degenerate, with the same signature as $g_{\mu\nu}$, as will be the case if we restrict to field configurations for which the theory is ``weakly coupled". By this, we mean that there exists a basis for which the components of the tensors $P_{gg}$, $P_{mg}$ and $P_{mm}$ are close to the corresponding expressions arising from the Einstein-Hilbert Lagrangian minimally coupled to conventional matter. 

Let $C = \det C_{\mu\nu}$ (in a RH basis). Explicitly this gives
\be
\label{Coverg}
 \frac{C}{g}=\frac{-\sqrt{-g}}{g} \epsilon^{\mu\nu\rho\sigma} C_{0\mu} C_{1\nu} C_{2\rho} C_{3\sigma} = -\frac{1}{4!} \epsilon^{\mu_1 \mu_2\mu_3\mu_4} \epsilon^{\nu_1 \nu_2\nu_3\nu_4} C_{\mu_1 \nu_1} C_{\mu_2\nu_2} C_{\mu_3 \nu_3}C_{\mu_4 \nu_4}  \ee 
We also have
\be
\epsilon^{\mu\nu\rho\sigma} = \left(\frac{C}{g}\right)^{1/2} \epsilon_{C}^{\mu\nu\rho\sigma}
\ee
where $\epsilon_C^{\mu\nu\rho\sigma}$ is defined by taking the volume form of $C_{\mu\nu}$ and raising indices with $(C^{-1})^{\mu\nu}$ (the inverse of $C_{\mu\nu}$). Hence \eqref{Cdef} implies
\be
C^{\mu_1 \mu_2 \mu_3 \, \nu_1 \nu_2 \nu_3} =\frac{C}{g} \left( -\frac{1}{2} \epsilon_C^{\mu_1 \mu_2 \mu_3 \rho}\epsilon_C^{\nu_1 \nu_2 \nu_3 \sigma} C_{\rho\sigma} \right)
\ee
Now note that the expression in parentheses can be obtained from the expression \eqref{Cdelta2} for $C^{\mu_1 \mu_2 \mu_3 \, \nu_1 \nu_2 \nu_3}$ arising from the {\it Einstein-Hilbert} Lagrangian simply by making the substitution $g_{\mu\nu} \rightarrow C_{\mu\nu}$ (and so $g^{\mu\nu} \rightarrow (C^{-1})^{\mu\nu}$). Hence by making this substitution in the Einstein-Hilbert result \eqref{EH1} (or by explicit computation) we obtain the following result for a general theory:
\ba
\label{Pgg_simp}
 C_{\mu \alpha} C_{\nu \beta} P_{gg}^{\alpha \beta \rho\sigma}(\xi) t_{\rho\sigma} &=&  \frac{C}{g} \left( -\frac{1}{2}(C^{-1})^{\rho\sigma}\xi_\rho \xi_\sigma \tilde{t}_{\mu\nu} + \xi_{(\mu} \tilde{t}_{\nu)\rho} (C^{-1})^{\rho\sigma} \xi_\sigma  \right. \nonumber \\ &{}& - \left. \frac{1}{2} C_{\mu\nu} \tilde{t}_{\rho\sigma} (C^{-1})^{\rho \alpha} (C^{-1})^{\sigma\beta} \xi_\alpha \xi_\beta  \right)
\ea
where
\be
\label{ttilde}
\tilde{t}_{\mu\nu} = t_{\mu\nu} - \frac{1}{2} C_{\mu\nu} (C^{-1})^{\rho\sigma} t_{\rho\sigma}
\ee

\subsection{Three dimensions}

A similar simplification occurs for $d=3$. In this case, the symmetries of $C^{\mu_1 \mu_2 \mu_3 \, \nu_1 \nu_2 \nu_3}$ imply that we can define a scalar $C$ by
\be
 C^{\mu_1 \mu_2 \mu_3 \, \nu_1 \nu_2 \nu_3} =-\frac{C}{2} \epsilon^{\mu_1 \mu_2 \mu_3}\epsilon^{\nu_1 \nu_2 \nu_3} 
\ee
hence
\be
 C_{\mu\rho\alpha}{}^{\nu\sigma\beta}=  \frac{C}{2}\delta_{\mu\rho \alpha}^{\nu\sigma \beta} 
\ee
so comparing with \eqref{Cdelta}, we see that the Einstein-Hilbert action gives $C=1$ and, for a general theory, $P_{gg}^{\mu\nu\rho\sigma\alpha\beta}$ can be obtained from the corresponding expression for the Einstein-Hilbert action simply by multiplying by $C$. In general, the scalar $C$ depends on the background fields and their first and second derivatives. 

\subsection{Five dimensions}

For $d=5$ the symmetries \eqref{Csymall} imply that we can write
\be
 C^{\mu_1 \mu_2 \mu_3 \, \nu_1 \nu_2 \nu_3} = \frac{1}{4} \epsilon^{\mu_1 \mu_2 \mu_3 \alpha_1 \alpha_2} \epsilon^{\nu_1 \nu_2 \nu_3 \beta_1 \beta_2} N_{\alpha_1 \alpha_2 \beta_1 \beta_2}
\ee
where $N_{\mu\nu\rho\sigma}=N_{[\mu \nu]\rho\sigma}=N_{\mu\nu [\rho\sigma]}$ and $N_{\mu\nu\rho\sigma} = N_{\rho\sigma \mu\nu}$. 
The symmetries \eqref{Csymbianchi} reduce to $N_{\mu[\nu\rho\sigma]}=0$ so $N_{\mu\nu\rho\sigma}$ has the symmetries of a metric-derived Riemann tensor. (For a Lovelock theory, the contribution of the $p=2$ term in \eqref{lovelock} to $N_{\mu\nu\rho\sigma}$ is proportional to $R_{\mu\nu\rho\sigma}$.)

\section{Characteristics}

\label{sec:chars}

\subsection{Definition of physical characteristics}

The usual definition states that a (real) covector $\xi_\mu$ is characteristic iff there exists a non-zero vector $T \equiv (t_{\mu\nu},t_I)$ such that ${\cal P}(\xi) T=0$, i.e., 
\begin{subequations}
\be
\label{char_eqa}
P_{gg}^{\mu\nu\rho\sigma}(\xi) t_{\rho\sigma} + P_{mg}^{I \mu\nu }(\xi) t_I = 0
\ee
and
\be
\label{char_eqb}
P_{mg}^{I\mu\nu}(\xi)t_{\mu\nu} + P_{mm}^{IJ}(\xi)t_J=0
\ee
\end{subequations}
where we have made use of the symmetry of ${\cal P}(\xi)$. However, this definition is not appropriate in theories with a gauge symmetry such as the theories we are studying. The reason is that \eqref{Psym2} implies that, for any $\xi_\mu$, taking $t_{\mu\nu} = \xi_{(\mu} X_{\nu)}$ and $t_I=0$ gives (for any $X_\mu$) a solution of the above equations. This is a consequence of the diffeomorphism invariance of the theory: such $t_{\mu\nu}$ corresponds to a ``high frequency gauge transformation". We can deal with this by following the approach of \cite{christodoulou2008mathematical} and working with gauge equivalence classes. So define an equivalence relation $t'_{\mu\nu} \sim t_{\mu\nu}$ if $t'_{\mu\nu} = t_{\mu\nu} +  \xi_{(\mu} X_{\nu)}$ for some $X_\mu$. The equations above depend only on the equivalence class $[t_{\mu\nu}]$ to which $t_{\mu\nu}$ belongs. Hence we can regard ${\cal P}(\xi)$ as acting on vectors of the form $([t_{\mu\nu}],t_I)$. We say that a real, non-zero, covector $\xi_\mu$ is characteristic iff there exists non-zero $([t_{\mu\nu}],t_I)$ satisfying the above equations. 

If the ``matter" is a Maxwell field ($t_I \rightarrow t_\mu$) then there is an additional gauge symmetry arising from electromagnetic gauge transformations. In this case we define $t'_\mu \sim t_\mu + c \xi_\mu$ for any constant $c$, and regard ${\cal P}(\xi)$ as acting on vectors of the form $([t_{\mu\nu}],[t_\mu])$. So real, non-zero $\xi_\mu$ is characteristic if there exists non-zero $([t_{\mu\nu}],[t_\mu])$ satisfying the above equations. 

We will focus on $d=4$ ``weakly coupled" theories. By this, we mean that the fields are such that $P_{gg}$, $P_{mg}$ and $P_{mm}$ are small deformations of the corresponding results for a ``conventional" theory of Einstein gravity minimally coupled to matter. In particular, $P_{mg}$ is small since it vanishes for a conventional theory. Furthermore, $C_{\mu\nu}$ is close to $g_{\mu\nu}$ so $C_{\mu\nu}$ is invertible and has the same signature as $g_{\mu\nu}$.

\subsection{Characteristics in four dimensions}

\label{sec:4dchar}

For $d=4$ recall that we defined $C_{\mu\nu}$ in \eqref{Cdef}. As in section \ref{sec:4d} we assume that $C_{\mu\nu}$ is non-degenerate, with the same (Lorentzian) signature as $g_{\mu\nu}$, as will be the case for a weakly coupled theory. The analysis can be split into two cases. 

{\bf Case 1} is defined by $(C^{-1})^{\mu\nu} \xi_\mu \xi_\nu \ne 0$.  The coefficient of $\tilde{t}_{\mu\nu}$ in \eqref{Pgg_simp} is non-zero. Hence, when this is substituted into \eqref{char_eqa}, the tensorial structure of this equation implies that $\tilde{t}_{\mu\nu}$ must take the form
\be
 \tilde{t}_{\mu\nu} = \xi_{(\mu}X_{\nu)} + \alpha C_{\mu\nu} + \beta C_{\mu\rho} C_{\nu\sigma} P_{mg}^{I\rho\sigma}(\xi)t_I
\ee
for some covector $X_\mu$ and scalars $\alpha,\beta$. Substituting back into \eqref{char_eqa}, one can solve to determine $\alpha,\beta$ (using \eqref{Psym2}). One can then invert the relation between $\tilde{t}_{\mu\nu}$ and $t_{\mu\nu}$ with the result
\be
\label{tsol}
 t_{\mu\nu} = \xi_{(\mu} X_{\nu)} + \frac{g}{C} [(C^{-1})^{\alpha\beta} \xi_\alpha \xi_\beta]^{-1} \left( 2C_{\mu\rho} C_{\nu\sigma} -  C_{\mu\nu} C_{\rho\sigma} \right) P_{mg}^{I\rho\sigma}(\xi) t_I
\ee
The first term of \eqref{tsol} is ``pure gauge". Thus, in this case, the ``gravitational" components $[t_{\mu\nu}]$ of the polarization vector are fully determined by the ``matter" components $t_I$. Substituting the above expression into \eqref{char_eqb} and rearranging gives
\be
\label{char_eq2}
 Q^{IJ}(\xi) t_J = 0
\ee
where
\be
 Q^{IJ}(\xi) \equiv  \frac{C}{g} (C^{-1})^{\mu\nu} \xi_\mu \xi_\nu P_{mm}^{IJ}(\xi) +  \left(2 C_{\mu\rho} C_{\nu\sigma} - C_{\mu\nu} C_{\rho\sigma} \right) P_{mg}^{I \mu\nu}(\xi) P_{mg}^{J \rho\sigma}(\xi)
\ee
This is a $N \times N$ symmetric matrix whose elements are homogeneous quartic polynomials in $\xi_\mu$. Note that an alternative expression for $C/g$ is given in \eqref{Coverg}. 

We can rewrite this matrix in a form that will prove useful later. Equation \eqref{PCI} relates $P_{mg}$ to the tensor $C^{I\mu\nu\rho\sigma}$ with Riemann symmetries. Let's now decompose this tensor into its ``Weyl" and ``Ricci" parts, defined w.r.t. the metric $C_{\mu\nu}$:
\be
\label{WIdef}
 W^{I\mu\nu\rho\sigma} \equiv C^{I \mu\nu\rho\sigma} - (C^{-1})^{\mu[\rho} D^{|I| \sigma]\nu} + (C^{-1})^{\nu[\rho} D^{|I| \sigma]\mu} + \frac{1}{3} D^I (C^{-1})^{\mu[\rho} (C^{-1})^{\sigma]\nu}
\ee
where
\be
\label{Dricdef}
 D^{I\mu\nu} \equiv C_{\rho\sigma} C^{I \mu\rho\nu\sigma} \qquad \qquad D^I \equiv C_{\mu\nu} D^{I \mu\nu}
 \ee
 By definition, $W^{I\mu\nu\rho\sigma}$ has the same symmetries as $C^{I\mu\nu\rho\sigma}$ and is traceless in the sense that contracting any pair of indices with $C_{\mu\nu}$ gives a vanishing result. Writing $C^{I\mu\nu\rho\sigma}$ in terms of its Weyl and Ricci parts gives
 \be
  Q^{IJ}(\xi) = -(C^{-1})^{\mu\nu} \xi_\mu \xi_\nu F^{IJ\alpha \beta}\xi_\alpha \xi_\beta + 2 C_{\mu \alpha} C_{\nu\beta} W^{I\mu\rho\nu\sigma} \xi_\rho\xi_\sigma W^{J \alpha \gamma \beta\delta} \xi_\gamma \xi_\delta 
 \ee
where $F^{IJ\alpha \beta} \equiv F^{IJ(\alpha \beta)}\equiv F^{(IJ)\alpha \beta}$ is defined by
\ba
F^{IJ\gamma\delta}\xi_\gamma \xi_\delta &=&- \frac{C}{g} P_{mm}^{IJ \gamma \delta}\xi_\gamma \xi_\delta-2C_{\alpha \mu}C_{\beta\nu} D^{(I|\mu\nu|}W^{J) \alpha \gamma \beta \delta} \xi_\gamma \xi_\delta - \frac{1}{3} D^{(I} D^{J) \gamma\delta} \xi_\gamma\xi_\delta \nonumber \\
&+& C_{\alpha\beta} D^{I \alpha\gamma} \xi_\gamma D^{J \beta \delta} \xi_\delta-(C^{-1})^{\gamma\delta}\xi_\gamma \xi_\delta \left( \frac{1}{2} C_{\mu\alpha} C_{\nu\beta} D^{I \mu\nu} D^{J \alpha \beta} - \frac{1}{6} D^I D^J \right)
\ea

Let's first discuss the case where the $\phi_I$ are real scalar fields. For \eqref{char_eq2} to admit a non-trivial solution we need $Q(\xi)=0$ where 
\be
 Q(\xi) \equiv \det Q^{IJ}(\xi)
\ee 
is a homogeneous polynomial in $\xi_\mu$ of degree $4N$ (where $N$ is the number of scalar fields). Fix some basis $\{f^0,f^i \; (i=1,2,3)\}$ for the cotangent space and write $\xi = \xi_0 f^0 + \xi_i f^i$. If we fix (real) $\xi_i$ then $Q(\xi)=0$ is a polynomial equation for $\xi_0$ of degree $4N$, so there are at most $4N$ real roots. Associated with each such root is a polarization $t_I$ which uniquely determines $[t_{\mu\nu}]$ via \eqref{tsol}. Hence, for a given direction $\xi_i$ we obtain at most $4N$ distinct characteristics $(\xi_\mu,[t_{\mu\nu}],t_I)$. However, our system has $N+2$ degrees of freedom ($N$ scalar and $2$ graviton) so, as long as the equations are hyperbolic in character (and the basis is chosen appropriately), there should be $2N+4$ physical characteristics: for each degree of freedom there should be ``future" and ``past" directed characteristics with the prescribed $\xi_i$. If $N>2$ then $4N>2N+4$, which implies that some of the roots $\xi_0$ of $Q(\xi)$ must be complex (or repeated). On the other hand, if $N=1$ then $4N<2N+4$ so the (quartic) polynomial $Q(\xi)$ cannot describe all physical characteristics -- the ``missing" characteristics correspond to case 2 below. 
 
Now we discuss briefly the case where the ``matter" is a Maxwell field, i.e., we take $\phi_I = A_\mu$. We now have a gauge freedom so $t_\mu \propto \xi_\mu$ is a ``pure gauge" solution of \eqref{char_eq2}, i.e., for any $\xi_\mu$ we have $Q^{\mu\nu}(\xi) \xi_\nu=0$, so $Q^{\mu\nu}(\xi)$ has rank at most $3$. To identify the physical characteristics we require that $Q^{\mu\nu}$ has rank strictly less than $3$, which is equivalent to the vanishing of the subdeterminant 
\be
 \Delta_{\mu\nu} = \epsilon_{\mu \rho_1 \rho_2 \rho_3} \epsilon_{\nu \sigma_1 \sigma_2 \sigma_3} Q^{\rho_1 \sigma_1} (\xi) Q^{\rho_2 \sigma_2}(\xi) Q^{\rho_3 \sigma_3}(\xi)
\ee
This is a homogeneous polynomial of degree $12$. Using $\xi_\nu Q^{\mu\nu}(\xi)=0$ one can show (compare \cite{Balakin:2017eur})
\be
 \Delta_{\mu\nu} = \xi_\mu \xi_\nu \Delta(\xi)
\ee
where $\Delta(\xi)$ is a homogeneous polynomial of degree $10$. So characteristics must satifisfy $\Delta(\xi)=0$. Fixing $\xi_i$ as above, this equation admits $10$ roots for $\xi_0$. However there are only $4$ physical degrees of freedom (2 graviton and 2 photon), so $2$ roots of $\Delta$ must be complex (or repeated).

{\bf Case 2.} This is defined by $\xi_\mu$ being null w.r.t. $C_{\mu\nu}$:
\be
\label{nullxi}
 (C^{-1})^{\mu\nu}\xi_\mu \xi_\nu=0
\ee
Introduce a null (w.r.t. $C_{\mu\nu}$) tetrad such that
\be
\label{nullbasis}
\xi_\mu = \delta^0_\mu\qquad 
 C_{00} = C_{11} = C_{0\hat{i}} = C_{1\hat{i}}=0 \qquad C_{01} = -1 \qquad C_{\hat{i}\hat{j}} = \delta_{\hat{i}\hat{j}}
\ee
where indices $\hat{i},\hat{j}$ take values $2,3$. In such a basis, the ``pure gauge" components of $t_{\mu\nu}$ are $t_{0\mu}$. We need to solve \eqref{char_eqa} and \eqref{char_eqb}. Using \eqref{Pgg_simp}, \eqref{char_eqa} reduces to
\be
\label{ttrans}
 \frac{C}{2g} t_{\hat{i}\hat{i}} + P_{mg}^{I 11}(\xi) t_I=0 \qquad 
 \frac{C}{2g} t_{1\hat{i}} + P_{mg}^{I 1\hat{i}}(\xi) t_I = 0 \qquad
 \frac{C}{g} t_{11}  + P_{mg}^{I \hat{i}\hat{i}}(\xi) t_I = 0
\ee
\be
\label{tracesym}
 \left(  P_{mg}^{I \hat{i}\hat{j}}(\xi) - \frac{1}{2} \delta_{\hat{i}\hat{j}} P_{mg}^{I \hat{k}\hat{k}}(\xi)  \right) t_I = 0
\ee
Equations \eqref{ttrans} fix the (gauge-invariant) ``longitudinal part" of $t_{\mu\nu}$ in terms of $t_I$, i.e., they fix $(C^{-1})^{\mu\nu} \xi_\mu \tilde{t}_{\nu\rho}$ where $\tilde{t}_{\mu\nu}$ is given by \eqref{ttilde}. The traceless part of $t_{\hat{i}\hat{j}}$ is not restricted by the above equations: this part has two independent components, corresponding to the 2 graviton polarizations.

Equation  \eqref{tracesym}  can be simplified by using \eqref{PCI} and writing $C^{I\mu\nu\rho\sigma}$ in terms of its Weyl and Ricci parts defined in \eqref{WIdef} and \eqref{Dricdef}. The result is
\be
\label{tracesym2}
 W^{I \hat{i} 0 \hat{j} 0 } t_I = 0
\ee 
This equation has a simple geometrical interpretation: it states that $\xi_\mu$ is a {\it principal null direction} (PND) of the tensor $W^{I\mu\nu\rho\sigma} t_I$. Note that this tensor has Weyl symmetries (w.r.t. the metric $C_{\mu\nu}$). So we have shown that \eqref{tracesym} is equivalent to
\be
\label{PND}
 \xi_\mu \; {\rm is\;a\;PND\; of} \; W^{I\mu\nu\rho\sigma} t_I
\ee 
A non-zero Weyl tensor admits exactly $4$ (possibly degenerate) principal null directions (up to scaling $\xi_\mu \sim \lambda \xi_\mu$). Given $\xi_\mu$, the above equation constrains $t_I$ such that $\xi_\mu$ is a PND. 

View the LHS of \eqref{tracesym2} as a linear map from the $N$ dimensional space of vectors $t_I$ to the 2d space of $2\times 2$ traceless symmetric matrices. Let $r \in \{0,1,2\}$ be the rank of this map. Then there is a $N-r$ dimensional space of vectors $t_I$ satisfying \eqref{tracesym2}. Hence there is a $N-r+2$ dimensional space of vectors $([t_{\mu\nu}],t_I)$ satisfying \eqref{char_eqa}. Note that the $N=1$ case is special because $r \le N$ so $r=2$ cannot occur for $N=1$. 

We now substitute these results into \eqref{char_eqb}. The result is an equation of the form $M^{IJ} t_J = \ldots$ where the RHS depends (linearly) only on the traceless part of $t_{\hat{i} \hat{j}}$. This is a linear equation constraining $t_I$ and the traceless part of $t_{\hat{i} \hat{j}}$. Let $s \le N$ be the rank of this system. Then this equation imposes $s$ conditions on $t_J$ and the traceless part of $t_{\hat{i} \hat{j}}$. Hence the space of solutions to \eqref{char_eqa} and \eqref{char_eqb} has dimension at least $N-r+2-s=(N-s) + (2-r)$. 

Consider the case in which the matter is a set of $N$ scalar fields. For $N=1$ (i.e. a Horndeski theory) we have $r \le 1$ hence $(N-s)+(2-r) \ge 1$ so there always exists a non-trivial solution to \eqref{char_eqa} and \eqref{char_eqb}. However, if $N \ge 2$ then ``generically" (i.e. for a generic field configuration of a generic theory) we might expect $s=N$ and $r=2$, suggesting that there are no solutions to \eqref{char_eqa} and \eqref{char_eqb}, i.e., that case 2 does not arise. However, there are certainly non-generic theories for which case 2 does arise (e.g. Einstein gravity minimally coupled to $N$ scalar fields has $r=s=0$). 

In the case where the matter is a Maxwell field (indices $I,J \rightarrow \mu,\nu$) we have a $4-r$ dimensional set of vectors $t_\mu$ allowed by \eqref{char_eqa}, but one of these is pure gauge (proportional to $\xi_\mu$), leaving $3-r$ physical photon polarizations $[t_\mu]$, plus the $2$ physical graviton polarizations, for a total of $5-r$ physical polarizations $([t_{\mu\nu}],[t_\mu])$ satisfying \eqref{char_eqa}. The symmetries of the principal symbol following from invariance under Maxwell gauge transformations imply that $M^{\mu\nu}(\xi) \xi_\nu=0$ so $M^{\mu\nu}(\xi)$ has non-trivial kernel and therefore $s \le 3$. Equations \eqref{char_eqa} and \eqref{char_eqb} have a space of solutions $([t_{\mu\nu}],[t_\mu])$ of dimension at least $5-r-s$. For example, conventional Einstein-Maxwell theory has $r=0$, $s=1$. However, for a generic background of a generic theory we might expect $r=2$ and $s=3$ and so case 2 may not arise. 

\section{Horndeski theories}

\label{sec:horn}

\subsection{Effective field theory}

\label{sec:EFT}

Consider Einstein gravity minimally coupled to a scalar field. The Lagrangian is
\be
\label{2deriv}
 L = R + X - V(\Phi) 
\ee
where $V(\Phi)$ is an arbitrary potential and 
\be
\label{Xdef}
 X=-\frac{1}{2}(\nabla \Phi)^{2}
\ee
The Lagrangian contains terms with up to $2$ derivatives of the fields. In EFT we add to this Lagrangian all possible higher-derivative scalars constructed from the fields. One can use field redefinitions to simplify the higher derivative terms. In particular, for a parity-symmetric theory, one can arrange that, after neglecting terms with $6$ or more derivatives, we have \cite{Weinberg:2008hq}
\be
\label{4dST}
 L = R + X -V(\Phi) + \frac{1}{2}\alpha(\Phi) X^2 + \frac{1}{4} \beta(\Phi) L_{GB} 
\ee 
where $V,\alpha,\beta$ are arbitrary functions and
\be
 L_{GB} = \frac{1}{4} \delta^{\mu_1 \mu_2 \mu_3 \mu_4}_{\nu_1 \nu_2 \nu_3 \nu_4} R_{\mu_1 \mu_2}{}^{\nu_1\nu_2} R_{\mu_3 \mu_4}{}^{\nu_3 \nu_4}
\ee
is the Euler density of the Gauss-Bonnet invariant (recall \eqref{gendeltadef}). From an EFT perspective, there is no reason to prefer $L_{GB}$ over, say, the square of the Weyl tensor. However, as we emphasised in the introduction, the above theory is preferred because it has second order equations of motion and admits a well-posed initial value formulation, at least when the theory is {\it weakly coupled}, i.e. when the $4$-derivative contributions to the principal symbol are small compared to the $2$-derivative contribution \cite{Kovacs:2020pns,Kovacs:2020ywu}.

 
The principal symbol of the above theory was calculated in Ref. \cite{Papallo:2017ddx} for the case $\alpha=0$. It is straightforward to include $\alpha$ (which affects only $P_{mm}$). One can then read off the effective metric defined in section \ref{sec:4d}:
\be
\label{C_EFT}
 C_{\mu\nu} = g_{\mu\nu} - \beta'(\Phi) \nabla_\mu \nabla_\nu \Phi - \beta''(\Phi) \nabla_\mu \Phi \nabla_\nu \Phi
\ee  
Turning to $P_{mg}$, recall that this is related to $C^{I\mu\nu\rho\sigma}$ by \eqref{PCI}. Since we have only a single scalar field we can drop indices $I,J,\ldots$ from our equations so $C^{I\mu\nu\rho\sigma} \rightarrow C^{\mu\nu\rho\sigma}$. Using the results of Ref. \cite{Papallo:2017ddx} we obtain:
 \be
 \label{CdualR}
  C^{\mu\nu\rho\sigma} = -\beta'(\Phi) \tilde{R}^{\mu\nu\rho\sigma}
 \ee
 where $\tilde{R}$ is the dual Riemann tensor:
 \be
  \tilde{R}_{\mu\nu\rho\sigma} = \frac{1}{4} \epsilon_{\mu\nu\alpha_1 \alpha_2} \epsilon_{\rho\sigma \beta_1 \beta_2} R^{\alpha_1 \alpha_2 \beta_1 \beta_2}
 \ee
Finally a straightforward calculation gives
\be
\label{Pmm4dST}
 P_{mm}^{\mu\nu} = -(1 +  \alpha(\Phi) X) g^{\mu\nu} + \alpha(\Phi) \nabla^\mu \Phi \nabla^\nu \Phi 
\ee 
Now we can explain precisely what we mean by this theory being weakly coupled. We require that there exists a basis such that the contribution of the $4$-derivative terms to the principal symbol is small compared to the contribution of the $2$-derivative terms. In a basis that is orthonormal w.r.t. $g_{\mu\nu}$, the non-zero contributions of the $2$-derivative terms to \eqref{C_EFT} or \eqref{Pmm4dST} are $\pm 1$. So we say that the theory is weakly coupled if there exists an orthonormal basis such that the components of all terms involving $\alpha,\beta$ in \eqref{C_EFT}, \eqref{CdualR} and \eqref{Pmm4dST} are small compared to $1$. In particular all components of $C^{\mu\nu\rho\sigma}$ must be small compared to $1$. 

More informally, in EFT, we expect $\alpha$ and $\beta$ to be proportional to $\ell^2$ where $\ell$ is a UV length scale. If the metric and scalar field vary over some length scale ${\cal L}$ then the theory will be weakly coupled provided ${\cal L}/\ell \gg 1$. Note that the theory might be weakly coupled in some region of spacetime but strongly coupled in some other region. 
 
 \subsection{General Horndeski theories}
 
The Lagrangian of a Horndeski theory takes the form $L= \sum_{k=2}^5 L_k$ where
\ba
L_{2}&=& \GG_{2}(\Phi,X) \qquad L_{3}= \GG_{3}(\Phi,X) \square\Phi \qquad L_{4}= \GG_{4}(\Phi,X)R+\partial_{X}\GG_{4}(\Phi,X)\delta^{\mu\rho}_{\nu\sigma}\,\nabla_{\mu}\nabla^{\nu}\Phi \, \nabla_{\rho}\nabla^{\sigma}\Phi \nonumber \\
 	L_{5}&=& \GG_{5}(\Phi,X)G_{\mu\nu}\nabla^{\mu}\nabla^{\nu}\Phi-\frac{1}{6}\partial_{X}\GG_{5}(\Phi,X)\delta^{\mu\rho\alpha}_{\nu\sigma\beta}\,\nabla_{\mu}\nabla^{\nu}\Phi \, \nabla_{\rho}\nabla^{\sigma}\Phi \, \nabla_{\alpha}\nabla^{\beta}\Phi 
\ea
with $X$ defined in \eqref{Xdef}. The functions $\GG_k$ are arbitrary functions of $\Phi$  and $X$.

The principal symbol of a Horndeski theory is given in Appendix B of \cite{Papallo:2017ddx} (where our $P_{gg}$ is denoted $\delta \tilde{P}_{gg}$ etc.) From this we can read off the effective metric defined in section \ref{sec:4d}:\footnote{For the special case $\partial_X \GG_4 = {\rm constant}$, $\GG_5=0$, our $C_{\mu\nu}$ reduces to the effective metric defined in \cite{Hajian:2020dcq}.} 
\be
\label{C_horn}
 C_{\mu\nu} = \left(\GG_4 - 2X \partial_X \GG_4 + X \partial_\Phi \GG_5 \right) g_{\mu\nu}- \left( \partial_X \GG_4 - \partial_\Phi \GG_5 \right) \nabla_\mu \Phi \nabla_\nu \Phi + X \partial_X \GG_5 \nabla_\mu \nabla_\nu \Phi
\ee
Expressions for $P_{mg}^{\mu\nu}(\xi)$ and $P_{mm}(\xi)$ can also be read off from Appendix B of \cite{Papallo:2017ddx}. They are lengthy so we will not repeat them here.

There is some degeneracy between the coefficients $\GG_k$. Furthermore, field redefinitions (e.g. a conformal transformation $g_{\mu\nu} \rightarrow \Omega(\Phi)^2 g_{\mu\nu}$) can be used to adjust these coefficients. We will assume that these coefficients are {\it smooth} functions and eliminate (most of) the degeneracy by imposing $(\partial_X \GG_2)(\Phi,0)=\GG_4(\Phi,0) =1$ and $\GG_3(\Phi,0)=0$, which corresponds to the 2-derivative terms in the action taking the form \eqref{2deriv}. We say that the theory is weakly coupled if there exists an orthonormal basis for which the components of the principal symbol are close to those of a 2-derivative theory of the form \eqref{2deriv}. This will be true if the background fields are slowly varying compared to any length scale $\ell$ defined by the functions $\GG_k$. More precise conditions are given in \cite{Papallo:2017qvl}.\footnote{Note that \cite{Papallo:2017qvl} uses slightly different definitions of $\GG_2$ and $\GG_4$.}

\subsection{Characteristics}

Let's summarize the results of our analysis of the characteristics in section \ref{sec:4dchar}. There are two cases. In case 1, $(C^{-1})^{\mu\nu} \xi_\mu \xi_\nu\ne 0$ and $\xi_\mu$ must satisfy $Q(\xi)=0$ where
 \be
 \label{Qhorn}
  Q(\xi) = -(C^{-1})^{\mu\nu} \xi_\mu \xi_\nu F^{\alpha \beta}\xi_\alpha \xi_\beta + 2 C_{\mu \alpha} C_{\nu\beta} W^{\mu\rho\nu\sigma} \xi_\rho\xi_\sigma W^{\alpha \gamma \beta\delta} \xi_\gamma \xi_\delta 
 \ee
with (symmetric) $F^{\alpha \beta}$ defined by
\ba
F^{\gamma\delta}\xi_\gamma \xi_\delta &=&  -\frac{C}{g} P_{mm}^{\gamma \delta}\xi_\gamma \xi_\delta-2C_{\alpha \mu}C_{\beta\nu} D^{\mu\nu}W^{ \alpha \gamma \beta \delta} \xi_\gamma \xi_\delta - \frac{1}{3} D D^{ \gamma\delta} \xi_\gamma\xi_\delta \nonumber \\
&+& C_{\alpha\beta} D^{ \alpha\gamma} \xi_\gamma D^{ \beta \delta} \xi_\delta-(C^{-1})^{\gamma\delta}\xi_\gamma \xi_\delta \left( \frac{1}{2} C_{\mu\alpha} C_{\nu\beta} D^{ \mu\nu} D^{ \alpha \beta} - \frac{1}{6}D^2  \right)
\ea
To recap on the notation: $W^{\mu\nu\rho\sigma}$, $D^{\mu\nu}$ and $D$ are the ``Weyl tensor", ``Ricci tensor" and ``Ricci scalar" formed from $C^{\mu\nu\rho\sigma}$ using the metric $C_{\mu\nu}$. 

$Q(\xi)$ is a homogeneous quartic polynomial in $\xi_\mu$. To write down the polarization vector $([t_{\mu\nu}],t)$ it is convenient to define $\psi$ by $t=(C/g) (C^{-1})^{\alpha\beta} \xi_\alpha \xi_\beta \psi$. From \eqref{tsol} and \eqref{PCI} we have
\be
  t_{\mu\nu} = 2\left( C_{\mu\rho} C_{\nu\sigma} - \frac{1}{2} C_{\mu\nu} C_{\rho\sigma} \right) C^{\rho\alpha \sigma \beta} \xi_\alpha \xi_\beta \psi \qquad \qquad  t=\frac{C}{g} (C^{-1})^{\alpha\beta} \xi_\alpha \xi_\beta \psi
\ee
where $t_{\mu\nu}$ is defined only up to addition of a pure gauge term. 

In case 2, $\xi_\mu$ must satisfy $(C^{-1})^{\mu\nu} \xi_\mu \xi_\nu=0$ and we need to solve \eqref{ttrans} and \eqref{tracesym2} (with $t_I \rightarrow t$) and then substitute the results into \eqref{char_eqb}. We can solve equation \eqref{tracesym2} by setting $t=0$. Equations \eqref{ttrans} then assert that the longitudinal components of $t_{\mu\nu}$ vanish, i.e., $(C^{-1})^{\mu\nu} \xi_\mu \tilde{t}_{\nu\rho}=0$. This means that the graviton polarization is ``transverse", where the notion of tranversality is defined w.r.t. $C_{\mu\nu}$ rather than $g_{\mu\nu}$. The non-zero components of $[t_{\mu\nu}]$ are the two components of the traceless matrix $t_{\hat{i}\hat{j}}$ (in the basis of \eqref{nullbasis}). Equation \eqref{char_eqb} reduces to
\be
\label{case2horn}
 W^{\hat{i} 0 \hat{j} 0 }  t_{\hat{i}\hat{j}}=0
\ee
which is at most one condition on these two components, leaving at least a single graviton polarization. So case 2 always admits a solution with a ``purely gravitational" ($t=0$) polarization that is transverse (w.r.t. $C_{\mu\nu}$). 

In special cases, there may be additional polarizations in case 2. For this to happen either \eqref{case2horn} becomes trivial or there exists a solution of \eqref{tracesym2} with $t \ne 0$. Either of these possibilities is equivalent to the vanishing of $W^{\hat{i}0 \hat{j} 0}$, i.e., to $\xi_\mu$ being a PND of $W^{\mu\nu\rho\sigma}$. When this happens, \eqref{ttrans} specifies the longitudinal components of $t_{\mu\nu}$ in terms of $t$. Substituting this into \eqref{char_eqb}, the tracless part of $t_{\hat{i}\hat{j}}$ drops out (because it appears in the combination \eqref{case2horn}, which is trivial) and so all non-zero terms in \eqref{char_eqb} are proportional to $t$. Thus this equation reduces to an expression of the form $S(\xi) t=0$. An expression for $S$ will be derived below (equation \eqref{Sdef}). Generically we expect $S(\xi) \ne 0$ and so $t=0$. Hence the analysis is the same as before except now there exists a 2-dimensional space of solutions of \eqref{case2horn}. So if $\xi_\mu$ is a PND then there are two transverse ``purely gravitatonal" polarizations. If $S(\xi)=0$ then there exists a third independent polarization $([t_{\mu\nu}],t)$ where $t \ne 0$ and $[t_{\mu\nu}]$ is purely longitudinal. 

We can relate this to the quartic polynomial $Q(\xi)$ of \eqref{Qhorn}. Since  $(C^{-1})^{\mu\nu} \xi_\mu \xi_\nu=0$, the first term in $Q$ vanishes, and the second term is
\be
\label{C2pos}
2 C_{\mu \alpha} C_{\nu\beta} W^{\mu\rho\nu\sigma} \xi_\rho\xi_\sigma W^{\alpha \gamma \beta\delta} \xi_\gamma \xi_\delta = 2 W^{\hat{i} 0 \hat{j} 0} W^{\hat{i} 0 \hat{j} 0}
\ee
The RHS vanishes iff $\xi_\mu$ is a PND. Hence, a covector $\xi_\mu$ null w.r.t. $(C^{-1})^{\mu\nu}$ satisfies $Q(\xi)=0$ iff $\xi_\mu$ is a PND of $W^{\mu\nu\rho\sigma}$. So in case 2, $\xi_\mu$ will give rise to more than $1$ independent polarization iff it also satisfies the quartic equation of case 1. 

A non-zero Weyl tensor admits $4$ (possibly coincident) PNDs (up to scaling $\xi_\mu \sim \lambda \xi_\mu$). It follows that if $W^{\mu\nu\rho\sigma} \ne 0$ then there exist exactly $4$ (possibly degenerate) directions $\xi_\mu$ satisfying both $(C^{-1})^{\mu\nu} \xi_\mu\xi_\nu=0$ and $Q(\xi)=0$. For these special directions there is a 2d space of purely gravitational polarizations. If $W^{\mu\nu\rho\sigma} = 0$ then {\it all} directions satisfying $(C^{-1})^{\mu\nu} \xi_\mu\xi_\nu=0$ will also satisfy $Q(\xi)=0$ and for all such $\xi_\mu$ there exists a 2d space of purely gravitational polarizations. 

\subsection{The characteristic cone and slowness surface}

\label{char_slowness}

From the above analysis, it follows that a non-zero real covector $\xi_\mu$ is characteristic iff either $Q(\xi)=0$ or $(C^{-1})^{\mu\nu} \xi_\mu \xi_\nu=0$. Thus we can write the condition for $\xi_\mu$ to be characteristic as $p(\xi)=0$ where $p$ is defined in equation \eqref{p_fact}.
Clearly $p(\xi)$ is a homogeneous polynomial of degree $6$ which factorises into a product of a quadratic and quartic polynomial. Note that degree $6$ is the minimum degree required to describe a second order system with three degrees of freedom. 

Fix a point in spacetime. We define the {\it characteristic cone} in the contangent space at that point as the set of characteristic covectors $\xi_\mu$, i.e., the set of (real) solutions of $p(\xi)=0$. Clearly this cone is the union of the quadratic cone $ (C^{-1})^{\mu\nu} \xi_\mu \xi_\nu=0$ and the quartic cone $Q(\xi)=0$. Recall that weak coupling ensures that $C_{\mu\nu}$ has the same signature as $g_{\mu\nu}$. Thus the quadratic cone is simply the null cone of the Lorentzian (inverse) metric $(C^{-1})^{\mu\nu}$. 

To understand the nature of the quartic cone, consider first the case of the 2-derivative theory \eqref{2deriv}, for which $C_{\mu\nu} = g_{\mu\nu}$, $C^{\mu\nu\rho\sigma}=0$ and $P_{mm}(\xi) = -g^{\mu\nu} \xi_\mu \xi_\nu$. Hence for this theory we have 
\be
\label{Q2deriv}
Q(\xi) = -(g^{\mu\nu} \xi_\mu \xi_\nu)^2 \qquad \qquad p(\xi) =  -(g^{\mu\nu} \xi_\mu \xi_\nu)^3 \qquad \qquad 2 {\rm \; derivative \;theory}
\ee
So in this case the quartic cone, and the full characteristic cone, degenerate to the null cone of the metric. Hence, in this 2-derivative theory, a hypersuface is characteristic iff it is null, so causality is determined by the null cone of the physical metric.

There is a more complicated class of theories for which $Q(\xi)$ factorises into a product of quadratic polynomials. This is the class of theories for which $W^{\mu\nu\rho\sigma}$ vanishes for any background solution. We will refer to such a theory as a {\it factorised theory}. For such a theory we have
\be
 \label{Qfact}
  Q(\xi) =- (C^{-1})^{\mu\nu} \xi_\mu \xi_\nu F^{\rho\sigma} \xi_\rho \xi_\sigma \qquad p(\xi) = - [(C^{-1})^{\mu\nu} \xi_\mu \xi_\nu]^2 F^{\rho\sigma} \xi_\rho \xi_\sigma \qquad {\rm factorised \; theory}
\ee
In this case, the quartic cone (and also the full characteristic cone) is the union of two quadratic cones, i.e., the null cones of $(C^{-1})^{\mu\nu}$ and $F^{\mu\nu}$. Hence for this class of theories the quadratic cone is a subset of the quartic cone. 
An example of a factorised theory is \eqref{4dST} with constant $\beta$ (which implies that the final term in \eqref{4dST} is topological). From \eqref{CdualR} we see that this theory has $C^{\mu\nu\rho\sigma}=0$, and hence $W^{\mu\nu\rho\sigma}=0$, for any background solution. Another example is a  Horndeski theory with $\GG_4 = 1$, $\GG_5=0$. In this case, $F^{\mu\nu}$ coincides with the ``effective" metric discussed previously in \cite{Deffayet:2010qz}. In both of these examples we have $C_{\mu\nu} = g_{\mu\nu}$.\footnote{The expressions in Appendix B of \cite{Papallo:2017ddx} suggest that the conditions for a Horndeski theory to be a factorised theory are $\partial_X \GG_5=\partial_X \GG_4-\partial_\Phi \GG_5 = 0$. Under these conditions, \eqref{C_horn} implies that $C_{\mu\nu}$ is conformal to $g_{\mu\nu}$ and so the quadratic cone is the same as the null cone of $g^{\mu\nu}$.}
 
A non-factorised theory has $W^{\mu\nu\rho\sigma} \ne 0$ in a generic background solution. However, for such a theory, there are non-generic background solutions for which $W^{\mu\nu\rho\sigma} = 0$ and hence $Q(\xi)$ factorises as above in such special backgrounds. 
For example, this occurs when the background solution is a cosmological solution with FLRW symmetry. This symmetry implies that $W^{\mu\nu\rho\sigma}$ and $C_{\mu\nu}$ have FLRW symmetry. Tracelessness of $W^{\mu\nu\rho\sigma}$ then implies $W^{\mu\nu\rho\sigma}=0$.  So in a FLRW background, we have $Q(\xi) =   -(C^{-1})^{\mu\nu} \xi_\mu \xi_\nu F^{\alpha \beta}\xi_\alpha \xi_\beta$ i.e., the quartic cone is the union of the null cones of $(C^{-1})^{\mu\nu}$ and $F^{\mu\nu}$. Hence this is another case for which the quadratic cone is a subset of the quartic cone. A covector is characteristic iff it is null w.r.t.  either $(C^{-1})^{\mu\nu}$ or $F^{\mu\nu}$. In cosmological terminology, the former case describes (purely gravitational) tensor modes and the latter case describes the scalar mode. Our two effective metrics determine the 2-derivative terms in the equations for tensor and scalar perturbations derived in \cite{Kobayashi:2011nu}. 

Now we consider a general theory and a general background solution for which the theory is weakly coupled, at least in some region. We will show that the quartic cone has two sheets, and that the quadratic cone lies between (or on) these sheets.

At weak coupling, we can pick a basis so that the components of $C_{\mu\nu}$ are close to those of $g_{\mu\nu}$. Choose such a basis $\{f^0,f^{i}\}$ ($i=1,2,3$) for the cotangent space, which is orthonormal w.r.t. $(C^{-1})^{\mu\nu}$, i.e., $(C^{-1})^{\mu\nu} = {\rm diag}(-1,1,1,1)$ and hence $g^{\mu\nu} \approx {\rm diag}(-1,1,1,1)$. Given a time-orientation, we  choose $f^0$ so that the corresponding dual basis vector is future-directed. In such a basis, fix the spatial components $\xi_{i}$ of $\xi_\mu$ and regard $Q(\xi)=0$ as a quartic equation to determine $\xi_0$ in terms of $\xi_{i}$ (which is assumed non-zero). In the 2-derivative theory, the roots of this quartic are the roots of $g^{\mu\nu} \xi_\mu \xi_\nu=0$, which we write as $\xi_0^\pm\approx \mp \sqrt{\xi_{i} \xi_{i}}$. These two roots correspond to the two components of the null cone of $g^{\mu\nu}$ and each root has degeneracy $2$. Weak coupling implies that, in our basis, the coefficients of the polynomial $Q(\xi)$ are small deformations of the coefficients in \eqref{Q2deriv}. Since the roots of a polynomial depend continuously on these coefficients, it follows that the $4$ roots $\xi_0$ of the quartic can be divided into $2$ pairs according to whether they are deformations of $\xi_0^+$ or of $\xi_0^-$.  The polynomial has real coefficients so each pair is either real, or is a pair of complex conjugate roots. 

The case of complex conjugate roots can be excluded as follows. Notice that, for the 2-derivative theory, $Q(\xi)$ is negative everywhere except on the null cone of $g^{\mu\nu}$. So, viewed as a function of $\xi_0$, $Q(\xi)$ is negative everywhere except at $\xi_0 = \xi_0^\pm$ where it vanishes. Hence when we deform to a weakly coupled theory, $Q(\xi)$ will be negative everywhere except possibly near $\xi_0^\pm$. Now evaluate $Q(\xi)$ on the null cone of $(C^{-1})^{\mu\nu}$, i.e., for $\xi_0 = \mp\sqrt{\xi_{i} \xi_{i}}$. The first term of $Q(\xi)$ vanishes. From \eqref{C2pos} we see that the final term in $Q(\xi)$ is non-negative. Hence $Q(\xi) \ge 0$ for $\xi_0 = \mp \sqrt{\xi_{i} \xi_{i}}$. It follows that $Q(\xi)$ must have a pair of (possibly degenerate) real roots near each of $\xi_0 = \mp \sqrt{\xi_{i} \xi_{i}}$, which excludes the possibility of complex roots. We label these roots as $\xi_{\rm in}^\pm$ and $\xi_{\rm out}^\pm$ where $|\xi_{\rm out}^\pm|\le \sqrt{\xi_{i} \xi_{i}}\le |\xi_{\rm in}^\pm|$. 
Note that the roots $\xi_{\rm out}^\pm$ and $\xi_{\rm in}^\pm$ are homogeneous in $\xi_i$, with degree $1$. 

We now see that the quartic cone is the union of two cones: an ``inner" cone $\xi_0 = \xi_{\rm in}^\pm$, lying inside (or on) the null cone of $(C^{-1})^{\mu\nu}$ (given by $\xi_0 = \mp \sqrt{\xi_i\xi_i}$) and an ``outer" cone $\xi_0 = \xi_{\rm out}^\pm$ lying outside (or on) the null cone of $(C^{-1})^{\mu\nu}$. (In both cases these are double cones, with the $\pm$ superscript distinguishing the two components of the double cone.) Inside the inner cone and outside the outer cone we have $Q(\xi)<0$ and between the inner and outer cones we have $Q(\xi) \ge 0$. 

This establishes that, for a weakly coupled theory, the characteristic cone is the union of three (double) cones: the quadratic cone  and the ``inner" and ``outer" cones just discussed. For given $\xi_{i}$, the three cones are associated with three different physical polarizations. In geometric optics, the speed of propagation of these polarizations is determined by these three cones. From the previous section we know that the quadratic cone is associated with a purely gravitational polarization whereas, generically, the inner and outer cones correspond to mixtures of scalar field and gravitational polarizations. For a weakly coupled theory, all three cones are close to the null cone of $g^{\mu\nu}$. 

Fix a point $q$ in spacetime. We define the {\it G\aa rding cone} $\Gamma_q^\pm$ as the connected component of $\{ \xi_\mu: p(\xi) \ne 0\}$ that contains $\mp f^0$ (weak coupling implies $p(f_0) \ne 0$). $\Gamma_q^\pm$ are the two open regions contained inside the inner sheet of the quartic cone. The results established above imply that $p(\xi)$ is a {\it hyperbolic polynomial}\footnote{\label{fn:hyp_pol}A homogeneous polynomial $p(\xi)$ is said to be a hyperbolic polynomial w.r.t. a covector $n_\mu$ iff $p(n) \ne 0$ and, for any $\xi_\mu$, the polynomial $p(\xi-\lambda n)$ has only real roots $\lambda$.} w.r.t. $f^0$, which implies that $\Gamma_q^\pm$ are convex sets \cite{hormander}. In the 2-derivative theory \eqref{2deriv}, the G\aa rding double cone $\Gamma_q^+ \cup \Gamma_q^-$ is the set of covectors that are timelike w.r.t. $g^{\mu\nu}$. In general, it is the set of covectors that are ``timelike" w.r.t. the causal structure defined by the equations of motion. In future we will sometimes suppress the dependence on $q$ and write $\Gamma^\pm$ instead of $\Gamma_q^\pm$.

As an example of the importance of the G\aa rding cone, consider the initial value problem, with initial data specified on a hypersurface $\Sigma$. Then the initial value problem is well-posed in the formulation of \cite{Kovacs:2020pns,Kovacs:2020ywu} if the initial data is chosen so that the theory is weakly coupled on $\Sigma$, and $\Sigma$ is ``spacelike" in the sense that its normal covector $n_\mu$ belongs to the G\aa rding cone. 

To visualise the characteristic cone, it is convenient to fix the scaling freedom $\xi_\mu \sim \lambda \xi_\mu$ by setting $\xi_0=-1$. This corresponds to taking the intersection of the characteristic cone with the plane $\xi_0=-1$. This defines the {\it slowness surface}, a surface in $\mathbb{R}^3$ with coordinates $\xi_{i}$. This name comes from the literature on elastic waves, see e.g. \cite{duff1960cauchy}. In a homogeneous elastic solid one can consider plane waves proportional to $\exp (i \xi_\mu x^\mu)$ with $\xi_\mu = (-\omega,\xi_i)$ and define the phase velocity $v_p = |\omega|/\sqrt{\xi_i\xi_i}$. Taking $\omega=1$ we then have $\sqrt{\xi_i\xi_i} = 1/v_p$ so the distance from the origin to a point on the slowness surface is the reciprocal of the phase velocity. Thus the inner sheet of the slowness surface corresponds to the ``fastest" degree of freedom. In our case, the slowness surface is the union of a two-sheeted quartic surface and a quadratic surface (in the above basis, a unit sphere), with the quadratic surface lying between (or on) the sheets of the quartic surface. The G\aa rding cone corresponds to the region inside the inner sheet of the quartic surface. 

We will need to determine whether the characteristic cone (or slowness surface) admits {\it singular points}. Consider an algebraic surface defined by a polynomial equation $f(x,y,z,\ldots)=0$. A singular point is a point on the surface at which the gradient of $f$ vanishes. At a singular point, vanishing of the gradient of $f$ implies that it might not be possible to draw a tangent plane at that point. Instead one can draw a tangent {\it cone} defined by the vanishing of the first not-identically-zero term in the Taylor expansion of $f$ about the singular point.  

In our case, if $\xi_\mu$ is a non-zero singular point of the characteristic cone then any multiple of $\xi_\mu$ is also a singular point, so such points fill out straight lines on the cone. In other words, it is only the direction of $\xi_\mu$ that is important so we will sometimes refer to such $\xi_\mu$ as a {\it singular direction}. Singular directions of the characteristic cone are in 1-1 correspondence with singular points of the slowness surface. We will now argue that $\xi_\mu$ is a singular direction iff it lies on both the quadratic and quartic cones, i.e., the singular directions are straight lines where the quadratic cone touches the quartic cone. Correspondingly, singular points of the slowness surface are points at which the quadratic surface touches the quartic surface. 

To see this, using  \eqref{p_fact} one finds that $\partial p /\partial\xi_\mu$ vanishes iff either (a) both $(C^{-1})^{\mu\nu} \xi_\mu \xi_\nu$ and $Q(\xi)$ vanish, i.e., $\xi_\mu$ lies on both the quadratic and quartic cones, or (b) $\partial Q/\partial \xi_\mu=0$ (which implies $Q=0$), i.e., $\xi_\mu$ is a singular direction of the quartic cone. In fact (b) is a special case of (a). To see this, fix $\xi_i$ and view $Q$ as a quartic polynomial in $\xi_0$ as above. If the roots are non-degenerate then they are smooth functions of the coefficients of the polynomial, and hence depend smoothly on $\xi_i$. Writing $Q$ in factorised form in terms of these roots we see that $\partial Q/\partial \xi_0 \ne 0$ when the roots are non-degenerate. Hence, at a singular point of $Q$, the roots of the quartic must be degenerate. From the discussion above we saw that such degeneracy occurs only when $\xi_\mu$ lies on both the quartic and quadratic cone. Hence (b) is a special case of (a). So $\xi_\mu$ is a singular direction of the characteristic cone iff it lies on both the quadratic and quartic cones. 

\begin{figure}
\begin{center}
\includegraphics[width=5cm]{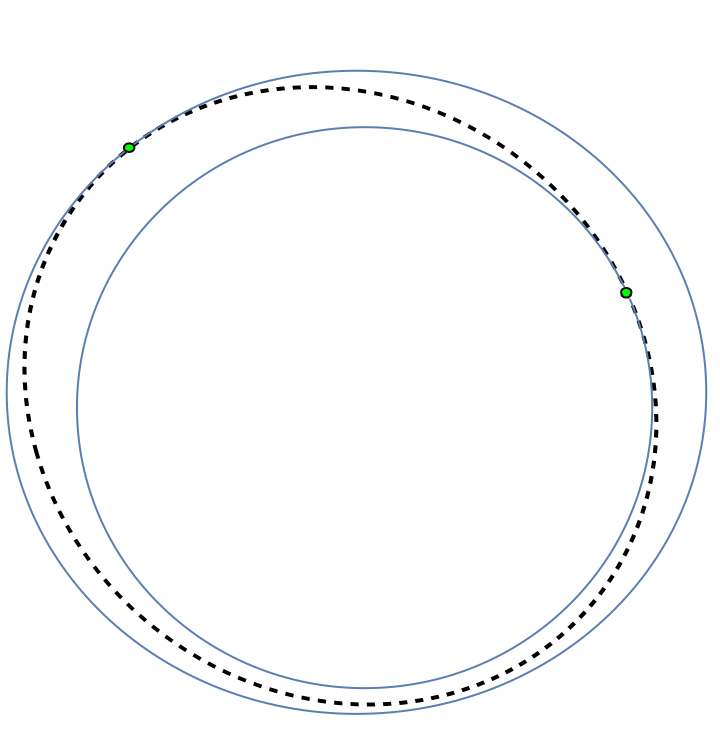}
\hspace{1.0cm}
\includegraphics[width=5cm]{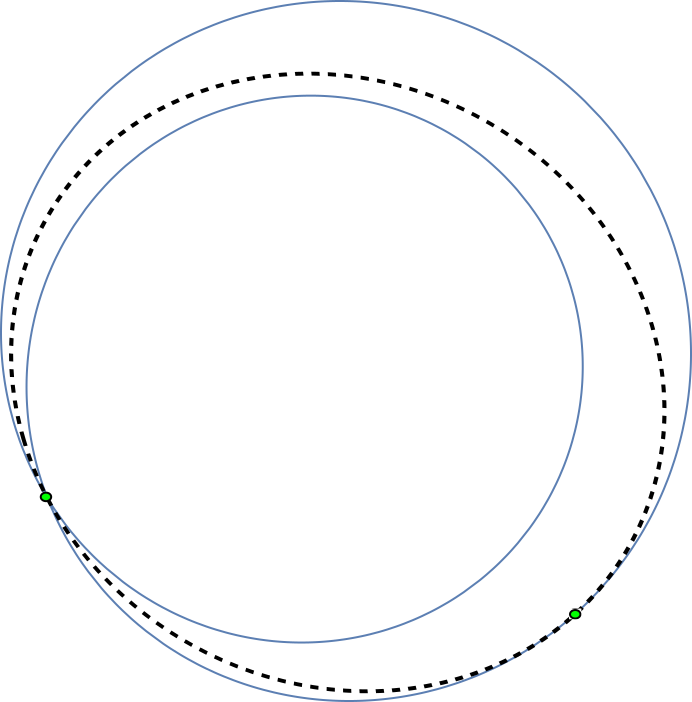}
\end{center}
\caption{Slowness surface when $W^{\mu\nu\rho\sigma} \ne 0$. The figures show the intersection of the slowness surface (in $\mathbb{R}^3$ with coordinates $\xi_i$) with a plane passing through the origin and two of the four points (the green dots) corresponding to principal null directions (PNDs). The dashed black curve shows the sphere corresponding to the null cone of $(C^{-1})^{\mu\nu}$. The solid blue curve corresponds to the quartic surface. The left sketch shows the generic behaviour, where the quartic surface is non-singular and the PNDs correspond to double points where the quadratic surface touches the quartic surface smoothly at the green points. The right sketch shows non-generic behaviour where one of the PNDs corresponds to a triple point at which the quartic surface is singular. (The figures are exaggerated for clarity; for a weakly coupled theory, all surfaces will be close to each other and to the ellipsoid corresponding to the null cone of the physical metric.)}
\label{fig:slowness}
\end{figure}

To summarise, a singular point on the slowness surface corresponds to a singular direction of the characteristic cone, along which the quadratic cone touches the quartic cone. The condition for (non-zero) $\xi_\mu$ to be such a direction is the vanishing of \eqref{C2pos}. A generic background solution of a non-factorised theory will have $W^{\mu\nu\rho\sigma} \ne 0$ and then \eqref{C2pos} vanishes for between $1$ and $4$ (generically $4$) distinct directions, corresponding to the (possibly coincident) principal null directions (PNDs) of $W^{\mu\nu\rho\sigma}$. This is shown in the left plot of Fig. \ref{fig:slowness}. On the other hand, for a factorised theory (or a non-generic background of a non-factorised theory) we have $W^{\mu\nu\rho\sigma} = 0$ and then the quadratic surface is a subset of the quartic surface (which sheet it coincides with may be different on different parts of the surface), so all points on the quadratic portion of the slowness surface are singular points. In either case, the analysis of the previous subsection shows that if $\xi_\mu$ is a singular direction then there exists a 2d space of ``purely gravitational" polarisations that (in geometric optics) can propagate in this direction. 

As an example, consider a spherically symmetric solution, i.e., both the metric and scalar field are spherically symmetric. Then $W^{\mu\nu\rho\sigma}$ will be spherically symmeric. Assume a non-factorised theory and that the background has $W^{\mu\nu\rho\sigma} \ne 0$ so there are $4$ (possibly degenerate) PNDs. A PND must be invariant under the spherical symmetry (as otherwise there would be a continuous family of PNDs) hence it must coincide with either the ``ingoing" or ``outgoing" radial null direction. Hence, from the result just established, any singular direction must coincide with one of these radial null directions. We can relate our discussion to the study of linear perturbations of static, spherically symmetric, solutions  in \cite{Kobayashi:2012kh,Kobayashi:2014wsa}. These perturbations can be classified by their parity. Odd parity perturbations are purely gravitational and reduce to a single ODE \cite{Kobayashi:2012kh}, with kinetic term determined by our effective metric $C_{\mu\nu}$. The even parity perturbations of \cite{Kobayashi:2014wsa} correspond to our quartic polynomial. Ref. \cite{Kobayashi:2014wsa} observed that one of the two ``radial velocities" for even parity perturbations coincides with the radial velocity of odd parity perturbations; this corresponds to our observation that the radial direction is a singular direction belonging to both the quadratic and quartic cones. 

In algebraic geometry, a {\it double point} is a singular point for which the Hessian $\partial^2 p/\partial \xi_\mu \partial \xi_\nu$ is non-vanishing. A {\it triple point} is a singular point for which this Hessian vanishes. At a singular point, this Hessian is proportional to $\xi_\rho (C^{-1})^{\rho(\mu} \partial Q/\partial \xi_{\nu)}$ so a triple point is a singular point at which $\partial Q/\partial \xi_\mu=0$, i.e., it is a singular point of the quartic cone.\footnote{
A triple point of the (degree $6$) slowness surface is a double point of the quartic surface but this is potentially confusing so we will not use the latter terminology.} Using this terminology, we can classify a singular direction of the characteristic cone as either a {\it double direction} or a {\it triple direction}, and we will refer to the corresponding points on the slowness surface as double points and triple points.  We saw above that if the roots of the quartic are non-degenerate then $\partial Q/\partial \xi_\mu \ne 0$. Hence, the roots of the quartic must be degenerate if $\xi_\mu$ is a triple direction, so the two sheets of the quartic cone coincide along such a direction. 
Since the quadratic cone lies between these sheets, it follows that all three sheets of the cone meet along a triple direction so this is a direction for which ``all three polarisations propagate at the same speed". 

A triple direction must satisfy the conditions for a singular direction, i.e., $\xi_\mu$ must be a PND of $W^{\mu\nu\rho\sigma}$. Evaluating $\partial Q/\partial \xi_\mu$ in the null basis of \eqref{nullbasis} and imposing the PND condition $W^{0\hat{i} 0\hat{j}}=0$ gives
\be
 \frac{\partial Q}{\partial \xi_\mu} = -2 (C^{-1})^{\mu\nu}\xi_\nu  F^{00}+ 8 W^{\mu 0 \hat{j} 0} W^{1 0 \hat{j} 0} = \delta^\mu_1 S
\ee 
where
\be
\label{Sdef}
 S \equiv 2 F^{00} + 8 W^{10\hat{j} 0}  W^{10\hat{j} 0} 
\ee
Thus $\xi_\mu$ is a triple direction iff $\xi_\mu$ is a PND which also satisfies $S=0$. Note that $S$ is independent of how the null basis vectors are chosen: $F^{00} = F^{\alpha \beta} \xi_\alpha \xi_\beta$ and it can be shown that the second term in $S$ is invariant under a change of basis when $\xi_\mu$ is a PND. 

For a double direction, $S \ne 0$ so $\partial Q/\partial \xi_\mu \propto \delta^\mu_1 = -(C^{-1})^{\mu\nu} \xi_\nu$ which shows that, along a double direction, the quartic cone and the quadratic cone have the same normal, i.e., they touch smoothly as shown in Fig. \ref{fig:slowness} (left plot). However, for a triple direction we have $\partial Q/\partial \xi_\mu=0$. Generically this means that one cannot draw a tangent plane to the slowness surface (or characteristic cone) at such a point, as shown in the right plot of Fig. \ref{fig:slowness}. Instead one has a tangent cone at such a point. However, in special cases this cone can degenerate to a plane, tangent to all three sheets of the slowness surface, and so in such cases one can draw a tangent. We will see below that this happens at a Killing horizon. 

If $W^{\mu\nu\rho\sigma} \ne 0$ then, generically, there is no reason to expect $S=0$ at one of the $4$ PNDs. So, generically, the PNDs correspond to double directions rather than triple directions. In a generic spacetime, the condition that $S=0$ at one of the PNDs defines a (maybe disconnected) hypersurface $\Sigma$ in spacetime. At a point $p \notin \Sigma$, the slowness surface will have (generically) $4$ double points. However, for $p \in \Sigma$ the slowness surface will have a triple point as well as double points. 

If $W^{\mu\nu\rho\sigma}=0$ (e.g. a factorised theory) then the condition for a triple direction reduces to $F^{\alpha \beta} \xi_\alpha \xi_\beta=0$, i.e., $\xi_\mu$ must be null w.r.t. both $(C^{-1})^{\mu\nu}$ and $F^{\mu\nu}$. We've already seen that when $W^{\mu\nu\rho\sigma}=0$, the quartic cone is the union of the null cones of these two metrics. The slowness surface is the union of the quadratic ellipsoids associated with each of these cones. All points on the ellipsoid defined by $(C^{-1})^{\mu\nu}$ are double points, and points that lie on both ellipsoids are triple points. If the ellipsoids intersect transversally then the lines of intersection are lines of triple points on the slowness surface. 

\subsection{Characteristic surfaces}

Consider linear perturbations around a background solution. In the high frequency (geometric optics) limit, a surface of constant phase is a characteristic surface. In the 2-derivative theory \eqref{2deriv}, a surface is characteristic iff it is null. This corresponds to the fact that, in this theory, high frequency disturbances propagate at the speed of light. For a more general Horndeski theory, characteristic hypersurfaces are generically non-null and, at high frequency, the $3$ physical degrees of freedom propagate with different speeds. For each speed, the above analysis above has determined a corresponding ``polarization eigenvector".

Given a solution arising from initial data specified on some surface $\Sigma$, if we disturb the data in a subregion $\Omega \subset \Sigma$ then the resulting disturbance will propagate into spacetime inside a region bounded by an ``outgoing" characteristic surface emanating from $\partial \Omega$ corresponding to the ``fastest" degree of freedom, i.e., a characteristic surface with normal $\xi_\mu \in \partial \Gamma^+$. Conversely, the {\it domain of dependence} of $\Omega$, the region of spacetime where the solution is uniquely determined by initial data on $\Omega$, will be the region inside the future and past-directed ingoing fastest characteristic hypersurfaces emanating from $\partial \Omega$ (see e.g. \cite{Reula:2004xd}  for results in this direction).

For the 2-derivative theory, a surfaces is characteristic iff it is null, and such surfaces are generated by null geodesics. In a more general theory, we have seen that $\xi_\mu$ is characteristic if it is null w.r.t. $C_{\mu\nu}$ hence a hypersurface that is null w.r.t. $C_{\mu\nu}$ is characteristic. Such surfaces are generated by null geodesics of $C_{\mu\nu}$. A hypersurface is also characteristic if its normal $\xi_\mu$ lies on the quartic cone. Such hypersurfaces are generated by {\it bicharacteristic curves} of the equation $Q(\xi)=0$. These are defined as follows \cite{courant2008methods}.  A bicharacteristic curve is a pair\footnote{
We sometimes won't distinguish between such a pair and its projection $x^\mu(t)$ to spacetime e.g. in the statement that null geodesics of $C_{\mu\nu}$ are bicharacteristic curves.} $(x^\mu(t),\xi_\nu(t))$ satisfying Hamilton's equations
\be
\label{bichar}
 \dot{x}^\mu = \frac{\partial Q}{\partial \xi_\mu} \qquad \qquad \dot{\xi}_\mu = - \frac{\partial Q}{\partial x^\mu}
\ee 
with initial condition $(x^\mu(0),\xi_\nu(0))$ chosen to satisfy $Q(x^\mu(0),\xi_\nu(0))=0$. Hamilton's equations then imply $Q(x^\mu(t),\xi_\nu(t))=0$, i.e, $\xi_\mu(t)$ is everywhere characteristic. One can also define bicharacteristic curves for the quadratic cone by replacing $Q$ with $(C^{-1})^{\mu\nu} \xi_\mu\xi_\nu$. These curves are simply the null geodesics of $C_{\mu\nu}$.  

The tangent vector to a bicharacteristic curve is a possible direction of propagation of a high frequency wavepacket. A characteristic covector $\xi_\mu$ can be regarded as the wavevector of a high-frequency plane wave, with $\xi_0$ fixed in terms of $\xi_i$ by solving the quartic equation, as discussed above. 
The group velocity can be defined as $c_g^i = -\partial \xi_0/\partial \xi_i$, viewed as a function of $\xi_i$.  Consider the tangent vector $X^\mu \equiv \partial Q/\partial \xi_\mu$ to a bicharacteristic curve associated with the quartic cone. Differentiating $Q(\xi)=0$ w.r.t. $\xi_i$ gives $-X^0 c_g^i + X^i=0$. Hence a bicharacteristic curve travels at the group velocity. Generically, for given $\xi_i$, we expect there to be three different group velocities associated with the 3 degrees of freedom, i.e., the three sheets of the characteristic cone. 

We argued above that, for a generic background of a non-factorised theory, the two sheets of the quartic cone do not meet except along triple directions which occur only on some (possibly disconnected) hypersurface $\Sigma$ in spacetime. For points on $\Sigma$ the slowness surface generically has an isolated triple point. Away from $\Sigma$, the two sheets of the quartic do not meet, so if $\xi_\mu(0) \in \partial \Gamma^\pm$ (the inner sheet) then $\xi_\mu(t) \in \partial \Gamma^\pm$ as long as $x^\mu(t)$ does not intersect $\Sigma$. Even when $x^\mu(t)$ does intersect $\Sigma$, the fact that the triple points are isolated points on the 2-dimensional slowness surface implies that a 2-parameter fine-tuning (of $\xi_\mu(0)$, which fixes $\dot{x}^\mu(0)$ via the first equation of \eqref{bichar}) would be required to hit a triple point. In other words, $\xi_\mu(0)$ must coincide with a particular point on the slowness surface in order for the bicharacteristic curve to hit the triple point when it reaches $\Sigma$. Thus, for a generic solution of a non-factorised theory, a bicharacteristic curve will not encounter a triple point of the quartic surface unless its initial direction is fine-tuned such that $\xi_\mu(0)$ coincides with certain isolated points on the slowness surface. 

What happens to a bicharacteristic curves that does reach a triple point of the slowness surface? There is now the possibility of the curve ``crossing" from the inner sheet to the outer sheet of the quartic surface, in which case it no longer corresponds to the fastest degree of freedom. However, when this happens, there will also be bicharacteristic curves which cross from the outer sheet to the inner sheet, so one can extend the original curve by gluing it to one of the latter curves. The resulting curve will be non-smooth at the point where the curves are glued. 

\subsection{Causal cone in the tangent space}

\label{sec:causalcone}

At any point $p$ of spacetime we define the {\it causal cone} ${\cal C}_p^\pm$ in the tangent space as the dual of the G\aa rding cone $\Gamma_p^\pm$:
\be
{\cal C}_p^\pm = \{ X^\mu : X^\mu \xi_\mu \le 0 \,\, \forall \,\, \xi_\mu \in \Gamma_p^\pm \}
\ee
The sets ${\cal C}_p^\pm$ are convex and closed. In the 2-derivative theory \eqref{2deriv}, ${\cal C}_p^+$ is the set of future-directed causal (or zero) vectors and ${\cal C}_p^-$ is the set of past-directed causal (or zero) vectors. In a more general theory, causal properties of the theory are not determined by the null cone of the metric but instead by the cones ${\cal C}_p^\pm$. Hence ${\cal C}_p^\pm$ should be used to define basic notions of causality. 

The strongest justification for this statement comes from results for linear hyperbolic (systems of) PDEs on $\mathbb{R}^d$ with constant coefficients. For such equations one can define $\Gamma^\pm$ in essentially the same way as we did above. With constant coefficients, one can identify the tangent space with spacetime in the same way that one does in special relativity, i.e., a vector in the tangent space corresponds to the position vector of a point in spacetime relative to some origin. We can now regard ${\cal C}_p^+$ (which, for constant coefficients, does not depend on $p$) as a cone in spacetime. It can be shown that this cone is the smallest closed convex cone containing the support of the retarded Green function with delta function source at the origin \cite{atiyah1970lacunas}\cite{hormander}. The significance of convexity is that if one can send a signal from the origin to points $X^\mu$ and $Y^\mu$ then one can also send a signal to $\lambda X^\mu$ and $(1-\lambda) Y^\mu$ for $\lambda \in [0,1]$. Hence one can first send a signal to the point with position vector $\lambda X^\mu$, and then from there one can send a signal to $\lambda X^\mu+(1-\lambda) Y^\mu$. So the region that one can send a signal to must be convex. 

Another example is given by Christodoulou's notion of a ``regularly hyperbolic" PDE, which encompasses equations with non-constant coefficients, and nonlinear equations, such as a perfect fluid, or an elastic solid \cite{action_principle}.\footnote{
This class of equations consists of theories admitting an action principle and such that high-frequency linearized perturbations have positive energy density. Unfortunately this class does not include gravitational theories, for which positivity of energy is more subtle.} For this class of equations, results analogous to \eqref{cone_vec} below imply that causality is determined by the cones ${\cal C}_p^\pm$.

We now consider the cones ${\cal C}_p^\pm$ for the theories discussed above. Consider first the case for which triple directions are absent, so the sheets of the quartic do not intersect each other (except at the origin). We now review an argument (see e.g.  \cite{action_principle}) which relates $\partial {\cal C}_p^\pm$ (the boundary of ${\cal C}_p^\pm$) to the set of tangent vectors to bicharacteristic curves. The set $\partial \Gamma_p^\pm$ is the innermost sheet of the quartic surface, which is non-singular in the absence of triple points (except at the origin). Now any non-zero $X^\mu \in {\cal C}_p^\pm$ defines a {\it supporting hyperplane} in the cotangent space, i.e., a plane through the origin $X^\mu \xi_\mu=0$ such that $\Gamma_p^\pm$ lies entirely on one side of this plane. If one considers how this plane varies as $X^\mu$ varies then it is clear that $X^\mu$ belongs to the boundary $\partial {\cal C}_p^\pm$ when this plane is a tangent plane to $\partial \Gamma_p^\pm$. But since $\partial\Gamma_p^\pm$ corresponds to a sheet of the quartic cone, it has normal $\partial Q/\partial \xi_\mu$. Hence if $X^\mu \in \partial C_p^\pm$ then $X^\mu \propto \partial Q/\partial \xi_\mu$ evaluated at some $\xi^\mu \in \partial \Gamma_p^\pm$. To fix the sign of the constant of proportionality, consider $\xi_\mu \in \partial \Gamma_p^\pm$ and pick $\delta \xi_\mu$ such that $\xi_\mu + \delta \xi_\mu \in \Gamma_p^\pm$. We argued above that $Q<0$ in $\Gamma_p^\pm$ in a weakly coupled theory. Hence $0>Q(\xi+\delta \xi) = \delta \xi_\mu \partial Q/\partial \xi_\mu + \ldots$ so $X^\mu$ must be a positive multiple of $\partial Q/\partial \xi_\mu$. Since $\partial Q/\partial x_\mu$ is a homogeneous cubic expression in $\xi_\mu$, the freedom to rescale $X^\mu$ by a positive constant just corresponds to the freedom to rescale $\xi_\mu$ by a positive constant, which is already present in the definition of $\partial \Gamma_p^\pm$. Hence we have shown that\footnote{Note that we allow $\xi_\mu$ to vanish on the RHS to recover the zero element of $\partial {\cal C}_p^\pm$.}
\be
\label{cone_vec}
 \partial {\cal C}_p^\pm = \{ \partial Q/\partial \xi_\mu : \xi_\mu \in \partial \Gamma_p^\pm \}\qquad \qquad {\rm (if \; no \; triple \; direction)}
\ee
However, from the previous section, a (non-zero) vector belongs to the set on the RHS iff it is tangent to a bicharacteristic curve associated with the innermost sheet of the quartic surface in the cotangent space. Since this innermost sheet corresponds to the ``fastest" degree of freedom, the associated bicharacteristic curves are the ``fastest" curves. Thus the boundary of ${\cal C}_p^\pm$ consists of tangent vectors to the fastest possible curves, which helps explain why ${\cal C}_p^\pm$ should be used to define notions of causality.

The RHS of \eqref{cone_vec} involves only the inner sheet of the quartic cone. The {\it wave cone} is the analogous set defined using all three sheets of the characteristic cone:
\be
 {\cal W}_p = \{ \partial Q/\partial \xi_\mu : Q(\xi)=0 \} \cup \{ X^\mu: C_{\mu\nu} X^\mu X^\nu=0\}
 \ee
 This can be divided into ``future" and ``past" cones ${\cal W}_p^\pm$. The wave cone also has $3$ sheets (each a double cone), with the cones $\partial {\cal C}_p^\pm$ corresponding to the outermost, i.e., ``fastest" sheet.\footnote{
Note that the fastest degree of freedom corresponds to the innermost sheet of the characteristic cone in the cotangent space but to the outermost sheet of the wave cone in the tangent space.} For PDEs with constant coefficients, the sheets of the wave cone correspond to the singular support of the Green function, i.e., to ``sharp" signals (see \cite{duff1960cauchy} for a clear discussion of this for the case of waves in an elastic solid). The three sheets of the wave cone are associated with the three different group velocities discussed above. 

To arrive at the result \eqref{cone_vec} we assumed that there are no triple directions. If there is a triple direction then the result \eqref{cone_vec} might not be true. In this case ${\cal C}_p^\pm$ is the {\it convex hull} of ${\cal W}_p^\pm$. The reason is that a triple point is a singular point of the quartic surface and, generically, one does not expect a tangent plane to exist at such a point. Instead there is a family of supporting hyperplanes, corresponding to a family of vectors $X^\mu$. 
If the slowness surface has an isolated triple point then this family of vectors fills out a planar section of the the convex hull, with normal $\xi_\mu$ (the triple direction). Thus if $\xi_\mu$ is a triple direction then it is associated with a continuous family of directions $X^\mu$, rather than with a unique direction as in \eqref{cone_vec}. 

As an example of this, consider a factorised theory. In such cases $Q(\xi)$ factorises into the product of two quadratic polynomials and ${\cal W}_p$ is the union of the null cones of the two effective metrics $C_{\mu\nu}$ and $(F^{-1})_{\mu\nu}$. If these null cones are concentric then ${\cal C}_p^\pm$ is simply the causal cone of the effective metric with the outermost null cone. However, if the cones of the two effective metrics have a non-trivial intersection then ${\cal C}_p^\pm$ is the convex hull of the union of these two cones. 

Now consider a non-factorised theory, with a background solution for which $W^{\mu\nu\rho\sigma} \ne 0$, and assume that the slowness surface has an isolated triple point. In this case, the quartic has an isolated singular point. This is similar to what happens for electromagnetism in an electrically anisotropic medium: a ``biaxial" crystal has a quartic slowness surface with $4$ singular points \cite{born2013principles}. As mentioned above, the singular points are associated with flat (planar) sections of the convex hull of ${\cal W}_p^\pm$, with a singular direction $\xi_\mu$ of the quartic surface corresponding to a family of vectors $X^\mu$ lying on such a planar section. In crystal optics, if one considers plane waves with wavevector $\xi_\mu$ then the associated $X^\mu$ corresponds to the direction of energy transport, i.e., to the direction of a ray. This leads to the phenomenon of conical refraction  \cite{born2013principles} in which a ray of electromagnetic waves with wavevector $\xi_i$ corresponding to a singular point on the slowness surface enters a crystal and is split into a family of rays, filling out a cone, with the associated $X^\mu$ lying on the planar section of $\partial {\cal C}_p^+$. This contrasts with what happens for generic (non-singular) $\xi_i$ where the incident ray is split into just two rays inside the crystal, corresponding to the two sheets of the slowness surface. 

In a generic background of a non-factorised theory, we have argued that the quartic surface is non-singular except on some hypersurface in spacetime. This hypersurface seems analogous to the case of a layer of a biaxial crystal of vanishing thickness, which is unlikely to lead to observable effects. However, there may be symmetrical (hence non-generic) solutions (perhaps stationary axisymmetric black hole spacetimes) for which the slowness surface admits a triple point {\it everywhere}. In such spacetimes, conical refraction may lead to interesting effects. 

\subsection{Causal structure and black holes}

\label{subsec:BHs}

I will use capital letters (e.g. ``Causal", ``Timelike") to distinguish notions defined w.r.t. ${\cal C}_p^\pm$ from the same notions defined in the traditional way using the metric. So we say a vector is future-directed Causal iff it is a non-zero element of ${\cal C}_p^+$ and future-directed Timelike iff it belongs to the interior of ${\cal C}_p^+$. Past directed Causal or Timelike vectors are defined by replacing ${\cal C}_p^+$ with ${\cal C}_p^-$. We say that a smooth curve is future-directed Causal (Timelike) iff its tangent vector is everywhere future-directed Causal (Timelike). Past directed Causal and Timelike curves are defined similarly. We define the Causal future of a set $W$ as the set ${\cal J}^+(W)$ consisting of points $p$ for which there exists a future-directed Causal curve from $W$ to $p$. Similarly the Chronological future ${\cal I}^+(W)$ is defined as the set of points $p$ for which there exists a future-directed Timelike curve from $W$ to $p$. The Causal and Chronological past are defined similarly. 

To define the notion of asymptotic flatness in the usual way, via conformal compactification, we need the spacetime curvature and scalar field derivatives to decay sufficiently rapidly that the causal structure near infinity is determined by the light cone of the metric in the usual way. More precisely, we need $C_{\mu\nu}$ to approach $g_{\mu\nu}$, and the quartic polynomial to approach \eqref{Q2deriv}, at appropriate rates at infinity. Then, near infinity, the causal structure of our theory will reduce to the causal structure defined using the metric in the usual way, which is preserved by conformal compactification. In the compactified spacetime we define the black hole region $\mathscr{B}$ of the spacetime manifold $M$ as $M \backslash {\cal J}^-(\mathscr{I}^+)$. The future event horizon $\mathscr{H}^+$ is defined as the boundary of $\mathscr{B}$. We expect this to be a ``fastest" outgoing characteristic hypersurface. In particular, when $\mathscr{H}^+$ is differentiable, its normal will belong to $\partial \Gamma^+$ (the inner sheet of the quartic). It would be very interesting to know whether this surface satisfies a version of the second law of black hole mechanics, i.e., is there some quantity that is non-decreasing along the bicharacteristic curves that generate this surface?

Now consider a stationary (i.e. time-independent) black hole solution. In a 2-derivative theory with suitable matter, it is known that the event horizon of such a black hole must be a {\it Killing horizon}. It would be interesting to try to prove such a result for theories of the type considered here, at weak coupling. Some evidence that such a result may exist, and some properties of a Killing horizon, are provided by the following proposition. 
\medskip

\noindent {\bf Proposition.} {\it Consider a smooth solution $(g_{\mu\nu},\Phi)$ which has a symmetry generated by a vector field $\xi^\mu$, i.e., $\xi^\mu$ is a Killing vector field and
\be
\label{dPhieq}
\xi^\mu \nabla_\mu \Phi=0
\ee 
Assume that this solution admits a Killing horizon ${\cal N}$ of $\xi^\mu$, i.e., ${\cal N}$ is a null hypersurface with normal $\xi_\mu$, and assume that the theory is weakly coupled on ${\cal N}$. Then ${\cal N}$ is a characteristic hypersurface and $\xi_\mu$ is a triple direction of the characteristic cone. Furthermore, the Hessian of $Q$ at that point is a non-zero multiple of $\xi^\mu \xi^\nu$. 
}

\medskip

This first part of this proposition is similar to previous results for Lovelock \cite{Izumi:2014loa,Reall2014} and Horndeski theories \cite{Minamitsuji:2015nca,Tanahashi:2017kgn}. In the Horndeski case, the interpretation of terms of the characteristic cone and the result for the Hessian are new. This proposition is proved in the Appendix. 

The significance of $\xi_\mu$ being a triple direction is that it implies that ${\cal N}$ is characteristic for all three physical polarizations. In particular, this implies that ${\cal N}$ is a fastest outgoing characteristic hypersurface (for a suitable definition of ``outgoing"). So this is consistent with the possibility that ${\cal N}$ is the event horizon of a stationary solution. 
 
To understand the significance of the above result for the Hessian, consider the quartic polynomial $Q(\omega)$ (we write $\omega_\mu$ here since we are using $\xi_\mu$ for the normal to our Killing horizon). Since $Q$ is a homogeneous quartic polynomial we have
\be
\label{QQ4}
 Q(\omega) = \frac{1}{4!} Q^{\mu\nu\rho\sigma} \omega_\mu\omega_\nu \omega_\rho \omega_\sigma
\ee
for some symmetric tensor $Q^{\mu\nu\rho\sigma}$ which can be read off from \eqref{Qhorn}. Since $\xi_\mu$ is a singular direction of the quartic, $\partial Q/\partial \omega_\mu$ vanishes for $\omega_\mu = \xi_\mu$ hence
\be
\label{Qxi3}
 Q^{\mu\nu\rho\sigma} \xi_\nu \xi_\rho \xi_\sigma=0
\ee
The Hessian of $Q$ at $\xi_\mu$ is
\be
 H^{\mu\nu} = \frac{1}{2} Q^{\mu\nu\rho\sigma} \xi_\rho \xi_\sigma
\ee
From \eqref{Qxi3} we have $H^{\mu\nu} \xi_\nu=0$ hence $H^{\mu\nu}$ is degenerate, with rank at most $3$. For a generic triple direction we expect that the rank will equal $3$. However, in the circumstances covered by our proposition we see that $H^{\mu\nu}$ has rank $1$, so this triple direction is non-generic. Consider the behaviour of $Q$ in a neighbourhood of $\xi_\mu$:
\be
 Q(\xi+\delta \xi) = \frac{1}{2} H^{\mu\nu} \delta \xi_\mu \delta\xi_\nu + \ldots
\ee
where the ellipsis denotes terms cubic or quartic in $\delta \xi_\mu$. Since $H^{\mu\nu} \propto \xi^\mu \xi^\nu$, the quadratic term vanishes if, and only if, $\xi^\mu \delta \xi_\mu=0$. This is the equation of a plane in the cotangent space with normal $\xi^\mu$. Hence in a neighbourhood of $\xi_\mu$, both sheets of the quartic cone degenerate to this plane with normal $\xi^\mu$ (which is also the normal to the quadratic cone as $(C^{-1})^{\mu\nu} \xi_\nu \propto \xi^\mu$ on ${\cal N}$: see \eqref{CA} in the Appendix). In particular, even though $\xi_\mu$ is a singular direction of the quartic cone, one can still define a unique tangent plane to the cone at this point. This has the following corollary:

\medskip

\noindent {\bf Corollary.} {\it If $p \in {\cal N}$ then a vector in ${\cal C}_p^+\cup {\cal C}_p^-$ is tangent to ${\cal N}$ if, and only if, it is a multiple of $\xi^\mu$.}

\medskip

\noindent {\it Proof}. Assume (by adjusting the sign if necessary) that $X^\mu \in {\cal C}_p^+$ and that $X^\mu$ is tangent to ${\cal N}$ so $X^\mu \xi_\mu=0$. We know that $\xi_\mu$ is characteristic and corresponds to a singular direction of the quartic cone. Hence $\xi_\mu$ belongs to both sheets of the quartic cone. In particular it belongs to the inner sheet $\partial \Gamma_p^+ \cup \partial \Gamma_p^-$. By choosing the appropriate sign we have $\pm \xi_\mu \in \partial \Gamma_p^+$. Consider the plane $X^\mu \omega_\mu=0$ in the cotangent space. We know that this plane contains $\pm \xi_\mu$. But, by definition of ${\cal C}_p^+$, this plane is a supporting hyperplane of $\Gamma_p^+$. So it is a supporting hyperplane that touches $\partial \Gamma_p^+$ (at $\pm \xi_\mu$). For a generic singular direction there could be many such supporting hyperplanes because a tangent plane is not defined at a generic singular point. However, 
we have just seen that our result for the Hessian implies that there {\it is} a unique tangent plane to $\partial \Gamma_p^+$ at $\pm \xi_\mu$, and this plane has normal $\xi^\mu$. Hence $X^\mu$ must be a multiple of $\xi^\mu$. Conversely, if $X^\mu \propto \xi^\mu$ then the plane $X^\mu \omega_\mu=0$ is tangent to $\partial \Gamma_p^+\cup \partial \Gamma_p^-$ at $\pm \xi_\mu$ and so either $\Gamma_p^+$ or $\Gamma_p^-$ lies in the region with $X^\mu \omega_\mu \le 0$ so $X^\mu \in {\cal C}_p^+\cup {\cal C}_p^-$. $\Box$

The point here is that, for a generic triple direction, we saw in section \ref{sec:causalcone} that the absence of a well-defined tangent plane implies that one might have e.g. a flat ``convex hull" section of $\partial {\cal C}_p^+$ (say) and this gives rise to the phenomenon of conical refraction. If $\xi_\mu$ had been a generic triple direction then conical refraction would occur within ${\cal N}$, i.e., from any point $p$ of ${\cal N}$ there would have been a cone (with narrow opening angle and containing $\xi^\mu$) of directions in which causal propagation tangential to ${\cal N}$ was possible.  But, in fact this does not happen because the triple direction is non-generic: the above result shows that causal propagation within ${\cal N}$ occurs only along the integral curves of $\xi^\mu$, i.e., the usual generators of ${\cal N}$. If the Killing horizon is axisymmetric as well as stationary then this means that the angular velocity can be defined in the usual way. 

Since $\xi_\mu$ is a singular direction, ${\cal N}$ is a null hypersurface w.r.t. $C_{\mu\nu}$ (as well as w.r.t. $g_{\mu\nu}$). Furthermore, since $\xi^\mu$ generates a symmetry of the solution, it follows that $\xi^\mu$ is a Killing vector field of $C_{\mu\nu}$ and hence ${\cal N}$ is a Killing horizon w.r.t. $C_{\mu\nu}$. What is the surface gravity of this Killing horizon? In the Appendix we prove:

\medskip
 \noindent {\bf Proposition.} {\it Under the same assumptions as the previous proposition, the surface gravity of ${\cal N}$ is constant (if ${\cal N}$ is connected) and the surface gravity of ${\cal N}$ w.r.t. $C_{\mu\nu}$ is the same as the surface gravity w.r.t. $g_{\mu\nu}$.}

\medskip

The first part of this proposition says that the horizon obeys the zeroth law of black hole mechanics. The second part of the proposition ensures that the Hawking temperature defined w.r.t. $C_{\mu\nu}$ is the same as that defined w.r.t. $g_{\mu\nu}$. If the solution admits a Euclidean section then there is a simpler way of seeing this: $C_{\mu\nu}$ is built from $g_{\mu\nu}$ and $\Phi$ so if these fields are smooth on the Euclidean section with a certain period for Euclidean time then $C_{\mu\nu}$ must also be smooth with the same choice of period. 

\subsection{Causality in effective field theory}

\label{subsec:EFTcausality}

As explained in the Introduction and in section \ref{sec:EFT}, some of the theories we have been considering can be motivated by EFT. In this section we will discuss briefly the question of whether the difference between the characteristic cone of the 4-derivative theory and the characteristic cone of the 2-derivative theory (i.e. the null cone of the metric) is actually observable in EFT. This is an issue that has been discussed several times in the literature, see e.g. \cite{Shore:2007um,Hollowood:2015elj,Goon:2016une}. A particularly detailed account has appeared recently \cite{deRham:2020zyh}. 

Consider the EFT of gravity coupled to a scalar field. As described in section \ref{sec:EFT}, the EFT action consists of the 2-derivative action \eqref{2deriv} supplemented by an infinite set of higher-derivative terms. The leading higher-derivative terms have $4$ derivatives and, after field redefinitions, can be written as in \eqref{4dST}. The coefficients  $\alpha$, $\beta$ of \eqref{4dST} are dimensionful with dimensions of length squared. In EFT these coefficients will be ${\cal O}(\ell^2)$ where $\ell$ is a length scale associated ``UV physics", e.g. the scale at which new massive fields start to play a role in the physics. Consider a field configuration that, in some coordinate chart, varies over a length scale $L$, i.e., derivatives of the fields are ${\cal O}(L^{-1})$. Validity of EFT requires $L/\ell \gg 1$; if this does not hold then one requires a full UV description of the physics. Then the 4-derivative terms in the equations of motion are suppressed relative to the 2-derivative terms by a factor or order $(\ell/L)^2$. Higher derivative terms are suppressed by higher powers of $(\ell/L)^2$. So in EFT, the 2-derivative theory provides the leading order description of the physics and the 4-derivative theory provides an improved description. The 4-derivative theory is weakly coupled, as we have assumed repeatedly above.\footnote{This does not imply that a solution of the 4-derivative theory must remain close to a solution of the 2-derivative theory. Secular effects, gradually accumulating over time, might cause a solution of the 4-derivative theory to diverge from a solution of the 2-derivative theory over a long enough time \cite{Flanagan:1996gw}. If this happens then the solution of the 4-derivative theory should provide the better description of the physics.}  

Since the 4-derivative theory should provide a better description of physics than the 2-derivative theory, one would expect that the characteristic cone of the 4-derivative theory, which we studied above, should provide a better description of causality than the characteristic cone of the 2-derivative theory (which is simply the null cone of the metric). However, as emphasised in \cite{deRham:2020zyh}, the difference between these two cones may not be observable in EFT. 

One way of seeing this is to consider how one might ``send a signal" from one point of spacetime to another. This can be done by using geometric optics to construct wavepackets which propagate along bicharacteristic curves. So consider a linear perturbation with wavelength $\lambda$, i.e., derivatives of the linearised fields are of order $\lambda^{-1}$. To apply the geometric optics approximation to the 4-derivative theory we assume that $\lambda$ is much shorter than any other length scale in the problem. One then finds that surfaces of constant phase are characteristic surfaces as defined above. However, the assumption that $\lambda$ is shorter than any other length scale is incompatible with the condition $\ell/\lambda \ll 1$ required for validity of EFT. 

Let's consider more carefully the size of different terms in the linearized equations. In the 2-derivative theory, the equation of motion gives us terms of order $\lambda^{-2}$, $L^{-1} \lambda^{-1}$ and $L^{-2}$, where $L$ is the scale over which the background solution varies (for example the linearised Einstein equation contains a term $R^{\mu\rho\nu\sigma} \delta g_{\rho \sigma}$ which is of order $L^{-2}$). In geometric optics we assume that first set of terms dominates, which requires $\lambda/L \ll 1$. 
When we include 4-derivative terms, the equation of motion now gives us additional terms of order $\lambda^{-2} (\ell/L)^2$, $L^{-1} \lambda^{-1} (\ell/L)^2$ and $L^{-2} (\ell/L)^2$ (there are no terms involving $\lambda^{-3}$ or $\lambda^{-4}$ because the equations of motion are second order). In applying geometric optics to the 4-derivative theory, we retain the terms of order $\lambda^{-2} (\ell/L)^2$. But $\lambda^{-2} (\ell/L)^2=L^{-2} (\ell/\lambda)^2 \ll L^{-2}$ because validity of EFT requires $\ell/\lambda \ll 1$. So within the regime of validity of EFT the terms of order $\lambda^{-2} (\ell/L)^2$ are negligible compared to the ``dispersive" 2-derivative terms of order $L^{-2}$ which are neglected in geometric optics. Thus, for consistency, we should also neglect the terms of order $\lambda^{-2} (\ell/L)^2$, in which case we retain just the terms of order $\lambda^{-2}$, which is just geometric optics of the 2-derivative theory (although applied to a background solution of the 4-derivative theory). Hence, within the regime of validity of EFT, geometric optics cannot distinguish between the characteristic cone of the 4-derivative theory and that of the 2-derivative theory. 

This raises the question of whether the analysis of this paper can tell us anything about EFT. The answer (probably) is yes, because the characteristic cone is relevant not just for geometric optics, but for many other properties of the 4-derivative theory. Since the (weakly coupled) 4-derivative theory makes sense as a self-contained classical theory, one might hope that some of the important theorems of General Relativity can be extended from the 2-derivative theory to the 4-derivative theory. The proofs of many of these theorems are based on causal properties of the theory. If these theorems can be extended to the 4-derivative theory then it seems likely that the characteristic cone, as defined in this paper, will provide the relevant notion of causality. So even if this notion of causality is not directly observable, it might provide instead a technical tool for establishing a result that does tell us something interesting in EFT. For example perhaps there exists an extension of the Penrose singularity theorem to the 4-derivative theory which involves a slightly modified definition of a trapped surface.\footnote{Such a theorem might assert that if there exists a trapped surface then either there exists an incomplete bicharacteristic curve or the solution becomes strongly coupled.} Or, as we have mentioned above, maybe there is an extension of the second law of black hole mechanics to these theories. 

\section{Discussion}

\label{sec:discuss}

The results of this paper suggest several opportunities for future research. I have considered in detail the class of theories  consisting of gravity coupled to a scalar field, with second order equations of motion. It would be interesting to perform a similar analysis for other theories with second order equations of motion. The class of theories of gravity coupled to an electromagnetic field will be discussed elsewhere. 

It would be interesting to determine the characteristic cone for some particular solutions of theories of the type \eqref{4dST} or more general Horndeski theories, for example black hole solutions. Stationary black hole solutions have been constructed numerically both in the spherically symmetric case (see e.g. \cite{Kanti:1995vq}) and in the rotating case (see e.g. \cite{Kleihaus:2011tg,Kleihaus:2015aje}). Our result on Killing horizons implies that, on the horizon of such a solution, the characteristic cone will admit a triple direction. But how does the cone behave in other directions? What happens for points not on the horizon? Also very interesting would be to study the characteristic cone for time-dependent solutions without symmetries, such as the solutions constructed numerically in \cite{East:2020hgw}. In particular one could study the properties of the event horizon, as defined above, of these dynamical solutions. 

Our definition of the characteristic polynomial provides a notion of (weak) hyperbolicity for Horndeski theories that is independent of any gauge-fixing procedure. The idea is that the characteristic polynomial $p(\xi)$ should be a {\it hyperbolic polynomial} (see footnote \ref{fn:hyp_pol}). We've seen that this is the case at weak coupling but it might fail for stronger coupling. Away from weak coupling, the effective metric $C_{\mu\nu}$ might not be invertible and so our definition of $p(\xi)$ can break down. However, this can be dealt with by defining 
\be
\tilde{p}(\xi) = (C/g) p(\xi) = {\cal C}^{\mu\nu} \xi_\mu \xi_\nu Q(\xi) \qquad \qquad {\cal C}^{\mu\nu} = -\frac{1}{3!} \epsilon^{\mu\rho_1\rho_2\rho_3} \epsilon^{\nu \sigma_1 \sigma_2 \sigma_3} C_{\rho_1 \sigma_1} C_{\rho_2 \sigma_2} C_{\rho_3 \sigma_3} 
\ee
which is always well-defined. The polynomials $p$ and $\tilde{p}$ have equivalent properties at weak coupling (when $C/g \approx 1$). It seems very unlikely that a well-posed formulation of the equations of motion will exist if $\tilde{p}(\xi)$ is not a hyperbolic polynomial. Conversely, if $\tilde{p}(\xi)$ is a hyperbolic polynomial for some generic class of backgrounds of interest then one might expect such a formulation to exist in these backgrounds. The formulation of \cite{Kovacs:2020pns,Kovacs:2020ywu} was proved to be well-posed at weak coupling. It would be interesting to know if it remains well-posed at strong coupling whenever $\tilde{p}(\xi)$ is a hyperbolic polynomial. It seems possible that this will be the case provided the auxiliary metrics $\tilde{g}^{\mu\nu}$ and $\hat{g}^{\mu\nu}$ of this formulation are chosen so that their null cones lie strictly outside the characteristic cone defined by $\tilde{p}(\xi)$. 

Ref. \cite{Ripley:2019hxt} considered spherically symmetric gravitational collapse in a theory of the form \eqref{4dST}. The spherically symmetric reduction of the theory was found to violate weak hyperbolicity when the fields become sufficiently strong. It is possible that a failure of hyperbolicity in the full theory occurs before the failure of hyperbolicity of the reduced theory. Computing the polynomial $\tilde{p}(\xi)$ for these backgrounds would provide a fairly simple way of determining whether or not this happens. 

\section*{Acknowledgments}

I am grateful to Aron Kovacs for helpful discussions and for comments on a draft. I am also grateful to Claudia de Rham and Andrew Tolley for a discussion. This work was supported by STFC grant no. ST/T000694/1.

\appendix
\section{Proof of propositions on Killing horizons}

\subsection{Proof of first proposition}

From \eqref{dPhieq}  we have
\be
\label{2dPhieq}
 0=\nabla_\mu (\xi^\nu \nabla_\nu \Phi) =  \xi^\nu \nabla_\mu \nabla_\nu \Phi +\nabla_\nu \Phi \nabla_\mu \xi^\nu
\ee 
contracting with $\xi^\mu$ gives
\be 
\xi^\mu\xi^\nu \nabla_\mu \nabla_\nu \Phi |_{\cal N}= -\kappa \xi^\nu \nabla_\nu \Phi =0 
 \ee
where $\kappa$ is the surface gravity of ${\cal N}$. Using this, and the fact that $\xi^\mu$ is null w.r.t. $g_{\mu\nu}$ we obtain from \eqref{C_EFT} or \eqref{C_horn} that $C_{\mu\nu} \xi^\mu \xi^\nu=0$ on ${\cal N}$, i.e., $\xi^\mu$ is null w.r.t. $C_{\mu\nu}$ on ${\cal N}$. Next we want to show that $\xi_\mu$ is null w.r.t. $(C^{-1})^{\mu\nu}$. Note that
\be
\label{nabla_xi}
\nabla_\mu \xi_\nu = \nabla_{[\mu} \xi_{\nu]} = 2\xi_{[\mu} \eta_{\nu]} 
\ee
for some $\eta_\mu$. The first equality is Killing's equation and the second equality, which holds only on ${\cal N}$, follows because $\xi_\mu$ is hypersurface orthogonal. Substituting into \eqref{2dPhieq} we have, on ${\cal N}$, 
\be
\label{2dPhieq2}
  \xi^\nu \nabla_\mu \nabla_\nu \Phi =-\xi_\mu  \eta^\nu \nabla_\nu \Phi 
\ee
Now, for either the theory \eqref{4dST} or for a general Horndeski theory we have
\be
\label{Cab}
 C_{\mu\nu} = a g_{\mu\nu} + b \nabla_\mu \nabla_\nu \Phi + c \nabla_\mu \Phi \nabla_\nu \Phi
\ee
for certain coefficients $a,b,c$. Hence, on ${\cal N}$, 
\be
\label{CA}
C_{\mu\nu} \xi^\nu = A \xi_\mu \qquad A \equiv a - b  \eta^\nu \nabla_\nu \Phi 
\ee
Non-degeneracy of $C_{\mu\nu}$ (for a weakly coupled theory) implies that the LHS is non-zero so $A \ne 0$. Rearranging we have $\xi^\mu = A(C^{-1})^{\mu\nu} \xi_\nu$ so contracting with $\xi_\mu$ we have, on ${\cal N}$, $0 = \xi_\mu \xi^\mu= A(C^{-1})^{\mu\nu} \xi_\mu \xi_\nu$. 
Hence, on ${\cal N}$, in a weakly coupled theory,
\be
 (C^{-1})^{\mu\nu} \xi_\mu \xi_\nu = 0 
\ee  
so the hypersurface ${\cal N}$ is null w.r.t. $C_{\mu\nu}$ as well as w.r.t. $g_{\mu\nu}$. It follows that ${\cal N}$ is a characteristic hypersurface associated with the quadratic cone. 

We will now show that $\xi_\mu$ also corresponds to a singular point of the quartic surface. We first show that certain components of the Riemann tensor vanish on ${\cal N}$. A Killing vector field satisfies 
\be
\label{killing_R}
\nabla_\mu \nabla_\nu \xi_\rho = R_{\rho\nu\mu\sigma} \xi^\sigma. 
\ee
Contract this equation with vectors $r^\rho$, $s^\nu$, $t^\mu$ that are tangent to ${\cal N}$ to obtain, on ${\cal N}$,
\be
\label{Riem_tang}
 R_{\rho\nu\mu\sigma} r^\rho s^\nu t^\mu \xi^\sigma =  r^\rho s^\nu t^\mu \nabla \mu \nabla_\nu \xi_\rho = r^\rho t^\mu \nabla_\mu ( s^\nu \nabla_\nu \xi_\rho) -  r^\rho( t^\mu \nabla_\mu s^\nu) \nabla_\nu \xi_\rho
\ee
On the RHS the first term involves a derivative w.r.t. $t^\mu$. Since $t^\mu$ is tangential to ${\cal N}$, we can use \eqref{nabla_xi} (which holds only on ${\cal N}$) to obtain
\ba
r^\rho t^\mu \nabla_\mu ( s^\nu \nabla_\nu \xi_\rho) &=& r^\rho t^\mu \nabla_\mu [s^\nu (\xi_\nu \eta_\rho - \xi_\rho \eta_\nu)] =- r^\rho t^\mu \nabla_\mu(s^\nu \eta_\nu \xi_\rho)=-s^\nu \eta_\nu r^\rho t^\mu\nabla_\mu \xi_\rho \nonumber \\
&=&-s^\nu \eta_\nu r^\rho t^\mu(\xi_\mu \eta_\rho - \xi_\rho \eta_\mu) = 0
\ea
where we have used the fact that $r^\mu,s^\mu$ and $t^\mu$ have vanishing contracting with $\xi_\mu$ because $\xi_\mu$ is normal to ${\cal N}$. The second term on the RHS of \eqref{Riem_tang} is
\ba
 r^\rho( t^\mu \nabla_\mu s^\nu) \nabla_\nu \xi_\rho&=& ( t^\mu \nabla_\mu s^\nu) r^\rho (\xi_\nu \eta_\rho - \xi_\rho \eta_\nu) = r^\rho \eta_\rho \xi_\nu t^\mu \nabla_\mu s^\nu = - r^\rho \eta_\rho s^\nu t^\mu \nabla_\mu \xi_\nu \nonumber \\
 &=& -r^\rho \eta_\rho s^\nu t^\mu ( \xi_\mu\eta_\nu - \xi_\nu \eta_\mu) = 0
\ea
we we have again used \eqref{Riem_tang} and the fact that $r^\mu,s^\mu$ and $t^\mu$ have vanishing contracting with $\xi_\mu$. Hence we have shown that, on ${\cal N}$,
\be
\label{Riem_zero}
 R_{\rho\nu\mu\sigma} r^\rho s^\nu t^\mu \xi^\sigma =0
\ee
for any vectors $r^\rho$, $s^\nu$, $t^\mu$ that are tangent to ${\cal N}$. 

We will now introduce a basis of vectors on ${\cal N}$ that is null w.r.t. $g_{\mu\nu}$.\footnote{This should not be confused with the basis of \eqref{nullbasis} that is null w.r.t. $C_{\mu\nu}$.} We choose $e_{1}^\mu \propto \xi^\mu$. We pick $e_{\hat{i}}^\mu$ ($\hat{i}=2,3$) to be a pair of orthonormal spacelike vectors that are tangent to ${\cal N}$ (and hence orthogonal to $\xi^\mu$). Finally we pick $e_{0}^\mu$ to be the unique null vector that is orthogonal to $e^\mu_{\hat{i}}$ and satisfies $g_{\mu\nu} e_{1}^\mu e_{0}^\nu=-1$. This defines a basis $\{e_{0}^\mu,e_{1}^\mu,e_{\hat{i}}^\mu, \hat{i}=2,3 \}$ such that the non-zero metric components are $g_{01}=g_{10} = -1$ and $g_{\hat{i}\hat{j}} = \delta_{\hat{i}\hat{j}}$. Note that the dual basis has $e^0_\mu  \propto \xi_\mu$.

Next we recall the concept of {\it boost weight}. A boost is a rescaling of the null basis vectors $e_{0} \rightarrow \lambda e_{0}$, $e_{1} \rightarrow \lambda^{-1} e_{1}$. A tensor component that scales by a factor of $\lambda^B$ has boost weight $B$. For example under this rescaling we have 
\be
\nabla_{1} \nabla_{\hat{i}} \Phi \equiv e_{1}^\mu e_{\hat{i}}^\nu \nabla_\mu\nabla_\nu \Phi \rightarrow \lambda^{-1}e_{1}^\mu e_{\hat{i}}^\nu \nabla_\mu \nabla_\nu \Phi 
\ee
hence the LHS has $B=-1$. The boost weight of a tensor component can be written as a sum where each subscript $0$ index contributes $+1$, each subscript $1$ index contributes $-1$, each superscript $0$ index contributes $-1$ and each superscript $1$ index contributes $+1$. Indices $\hat{i},\hat{j}$ do not contribute to $B$. So in the above example $B=-1+0=-1$. Note that boost weight is additive: if we consider the outer product of two tensors then the boost weight of a given component is the sum of the boost weights of the terms appearing in the product.

Our strategy now will be to show that, on ${\cal N}$, the negative boost weight components of all relevant tensors are zero. First consider \eqref{dPhieq} and \eqref{2dPhieq2}. In our basis these give
\be
\label{PhidPhibasis1}
 \nabla_{1} \Phi =0  \qquad \qquad \nabla_{1} \nabla_{1} \Phi = \nabla_{1} \nabla_{\hat{i}} \Phi =0
\ee 
hence the negative boost weight components of $\nabla_\mu \Phi$ and $\nabla_\mu \nabla_\nu \Phi$ vanish on ${\cal N}$. Note also that non-vanishing components of $g_{\mu\nu}$ have $B=0$. 

Taking $r=e_{\hat{i}}$, $s=e_{1}$, $t=e_{\hat{j}}$ or $r=e_{\hat{i}}$, $s=e_{\hat{j}}$, $t=e_{\hat{k}}$ in \eqref{Riem_zero} we now obtain
\be
\label{Rbasis1}
 R_{\hat{i}1\hat{j}1}=0 \qquad \qquad R_{\hat{i} \hat{j} \hat{k}1} =0
\ee
So far we have not used any equations of motion. We now consider the Einstein equation on ${\cal N}$ which can be written as
\be
G_{\mu\nu} = \ldots
\ee
where the RHS is a polynomial in $\nabla_\mu \Phi$, $\nabla_\mu \nabla_\nu \Phi$, $R_{\mu\nu\rho\sigma}$, $g_{\mu\nu}$ and $g^{\mu\nu}$ (with coefficients that are scalar functions of $\Phi$), where $R_{\mu\nu\rho\sigma}$ appears only with degree $0$ and $1$ (see Appendix A of \cite{Papallo:2017ddx}). Consider the negative boost weight components of this equation. The $B=-2$ component can be seen to be trivial so we consider the $B=-1$ component. The second equation of \eqref{Rbasis1} implies that the LHS is $R_{1\hat{i}} = -R_{{0}{1}{1} \hat{i}  }$. The negative boost weight components of all tensors on the RHS vanish, except possibly for $R_{\mu\nu\rho\sigma}$ which has possibly non-zero negative boost-weight components $R_{{0}{1}{1} \hat{i}  }$ (and those related by symmetry). Hence the Einstein equation reduces to 
\be
 -R_{{0}{1} {1} \hat{i}  }= X_{\hat{i} \hat{j}} R_{{0}{1}{1} \hat{j}  }
\ee
for some quantity $X_{\hat{i} \hat{j}}$ with $B=0$. Now in the 2-derivative theory \eqref{2deriv} the RHS of the Einstein equation does not involve the Riemann tensor hence $X_{\hat{i} \hat{j}}=0$ in this theory. Hence, in a general weakly coupled theory, $X_{\hat{i} \hat{j}}$ must be small and so the determinant of the above linear system is non-zero and the Einstein equation implies
\be
\label{R011i}
 R_{{0}{1}{1} \hat{i}  }=0
\ee
Combined with \eqref{Rbasis1} this shows that all negative boost weight components of the Riemann tensor vanish on ${\cal N}$. 

From \eqref{PhidPhibasis1} and \eqref{C_EFT} or \eqref{C_horn} we see that the negative boost weight components of $C_{\mu\nu}$ must all vanish on ${\cal N}$. Using $(C^{-1})^{\mu\nu} C_{\nu\rho} = \delta^\mu_\rho$ we obtain $(C^{-1})^{\mu 0} \propto \delta^\mu_1$ and hence negative boost weight components of $(C^{-1})^{\mu\nu}$ must also vanish on ${\cal N}$. 

Now we consider $C^{\mu\nu\rho\sigma}$. This is another polynomial in $\nabla_\mu \Phi$, $\nabla_\mu \nabla_\nu \Phi$, $R_{\mu\nu\rho\sigma}$, $g_{\mu\nu}$ and $g^{\mu\nu}$ with coefficients that are scalar functions of $\Phi$. This can be seen from \eqref{CdualR} (for the theory \eqref{4dST}) or from the expressions in Appendix B of \cite{Papallo:2017ddx}. We have shown that the negative boost weight components of all of these tensors vanish and so it follows that all negative boost weight components of $C^{\mu\nu\rho\sigma}$ must vanish on ${\cal N}$:
\be
\label{Ccpts}
 C^{{0}\hat{i}{0}\hat{j}}=0 \qquad  \qquad C^{{0}{1}{0} \hat{i}  }=0 \qquad C^{{0}\hat{i} \hat{j} \hat{k}} =0
\ee
Since all negative boost weight components of $C^{\mu\nu\rho\sigma}$ and $C_{\mu\nu}$ vanish, it follows that so must the negative boost weight components of $D^{\mu\nu}$ and $W^{\mu\nu\rho\sigma}$:
\be
 D^{00} = D^{0i} = 0 \qquad W^{0i0j} = W^{010i} = W^{0 ijk} = 0
\ee
We can write our quartic polynomial for a general covector $\omega_\mu$ as in \eqref{QQ4}. The tensor $Q^{\mu\nu\rho\sigma}$  is a polynomial in $D^{\mu\nu}$, $W^{\mu\nu\rho\sigma}$, $C_{\mu\nu}$, $(C^{-1})^{\mu\nu}$, $\nabla_\mu \Phi$, $\nabla_\mu \nabla_\nu \Phi$, $g_{\mu\nu}$ and $g^{\mu\nu}$. (Here we use the expression for $P_{mm}$ in \eqref{Pmm4dST} or the corresponding expression in Appendix B of \cite{Papallo:2017ddx}.) We have shown that negative boost weight components of all of these tensors vanish on ${\cal N}$. Hence the negative boost weight components of $Q^{\mu\nu\rho\sigma}$ must vanish on ${\cal N}$. In particular we have
\be
 Q(\xi) \propto Q^{0000} = 0
\ee
so $\xi_\mu$ lies on the quartic cone. Furthermore
\be
\left( \frac{\partial Q}{\partial \omega_\mu}\right)_{\omega=\xi} = \frac{1}{3!} Q^{\mu\nu\rho\sigma} \xi_\nu \xi_\rho \xi_\sigma \propto Q^{\mu000} = 0
\ee
hence $\xi_\mu$ is a singular direction of the quartic surface. Finally we have
\be
H^{\mu\nu} \equiv \left( \frac{\partial^2 Q}{\partial \omega_\mu\partial \omega_\nu}\right)_{\omega=\xi}  =\frac{1}{2!} Q^{\mu\nu\rho\sigma}  \xi_\rho \xi_\sigma \propto Q^{\mu\nu00} = \delta^\mu_1 \delta^\nu_1 Q^{1100}
\ee
For the 2-derivative theory \eqref{2deriv} we know that $Q$ takes the form \eqref{Q2deriv} for which $Q^{1100} < 0$. Hence, by continuity, we will have $Q^{1100} < 0$ in a weakly coupled theory. So we have shown that, at weak coupling, the Hessian of $Q$ at $\xi_\mu$ has rank $1$: $H^{\mu\nu} \propto \xi^\mu \xi^\nu$. 

\subsection{Proof of second proposition}

The surface gravity of ${\cal N}$ w.r.t. $g_{\mu\nu}$ is defined by
\be
\label{kappa2}
  \nabla_\mu \left( g_{\nu\rho} \xi^\nu \xi^\rho \right)|_{\cal N} = -2 \kappa \xi_\mu 
\ee
which is equivalent to
\be
\label{kappa_alt}
 \xi^\nu \nabla_\nu \xi^\mu |_{\cal N}= \kappa \xi^\mu
\ee
To prove the first part of the proposition (the zeroth law of black hole mechanics), contract \eqref{killing_R} with $\xi^\nu$ and rearrange to obtain
\be
R^\rho{}_{\nu\mu\sigma} \xi^\nu \xi^\sigma = \nabla_\mu \left( \xi^\nu \nabla_\nu \xi^\rho \right) - (\nabla_\mu \xi^\nu )\nabla_\nu \xi^\rho
\ee
Now let $t^\mu$ be tangent to ${\cal N}$. Contracting with $t^\mu$ and using \eqref{kappa_alt} gives, on ${\cal N}$
\ba
 R^\rho{}_{\nu\mu\sigma} \xi^\nu t^\mu \xi^\sigma&=& t^\mu \nabla_\mu \left(\kappa \xi^\rho \right) - t^\mu(\nabla_\mu \xi^\nu )\nabla_\nu \xi^\rho \nonumber \\
 &=& (t^\mu \nabla_\mu \kappa ) \xi^\rho + \kappa t^\mu \nabla_\mu \xi^\rho - t^\mu(\nabla_\mu \xi^\nu )\nabla_\nu \xi^\rho \nonumber \\
&=& (t^\mu \nabla_\mu \kappa ) \xi^\rho - \kappa t^\mu \eta_\mu \xi^\rho - t^\mu \eta_\mu \xi^\nu \eta_\nu \xi^\rho
\ea
using \eqref{nabla_xi} in the final line. Now contracting \eqref{nabla_xi} with $\xi^\mu$ and using \eqref{kappa_alt} gives
\be
\label{kappaetaxi}
 \kappa= -\eta^\mu \xi_\mu
\ee 
hence the final two terms cancel above, giving
\be
\label{R_kappa}
  R_{\rho\nu\mu\sigma} \xi^\nu t^\mu \xi^\sigma= (t^\mu \nabla_\mu \kappa ) \xi_\rho
\ee
Evaluating this equation in the null basis used in the proof of the previous proposition gives
\be
   (t^\mu \nabla_\mu \kappa )\delta_\rho^0\propto R_{\rho 1 \mu 1} t^{\mu} \qquad \Rightarrow \qquad  t^\mu \nabla_\mu \kappa  \propto R_{0 1 \hat{i} 1} t^{\hat{i}} 
\ee
using $t^0=0$ as $t^\mu$ is tangent to ${\cal N}$. The RHS vanishes because of  \eqref{R011i}. Hence $\kappa$ is constant on each connected component of ${\cal N}$. 

Now consider the surface gravity $\kappa_C$ of ${\cal N}$ defined w.r.t. $C_{\mu\nu}$:
\be
\label{kappaC}
  \nabla_\mu \left( C_{\nu\rho} \xi^\nu  \xi^\rho \right)|_{\cal N} = -2 \kappa_C C_{\mu\nu} \xi^\nu
\ee
We want to show $\kappa_C = \kappa$. 

Using \eqref{dPhieq}, \eqref{2dPhieq} and \eqref{Cab} we have
\be
  C_{\nu\rho} \xi^\nu  \xi^\rho = a g_{\nu\rho} \xi^\nu \xi^\rho - b (\xi^\nu \nabla_\nu \xi^\rho) \nabla_\rho \Phi
\ee
Hence on ${\cal N}$ using \eqref{kappa_alt} gives
\ba
  \nabla_\mu \left( C_{\nu\rho} \xi^\nu  \xi^\rho \right) &=& a  \nabla_\mu \left( g_{\nu\rho} \xi^\nu \xi^\rho \right) - \kappa \xi^\rho \left( \nabla_\mu b \nabla_\rho \Phi + b \nabla_\mu \nabla_\rho \Phi \right) -b \nabla_\mu ( \xi^\nu \nabla_\nu \xi^\rho) \nabla_\rho \Phi \nonumber \\
  &=& -2 a \kappa \xi_\mu + b \kappa \nabla_\mu \xi^\rho \nabla_\rho \Phi-b( \xi^\nu \nabla_\mu \nabla_\nu \xi^\rho) \nabla_\rho \Phi - b( \nabla_\mu \xi^\nu)(\nabla_\nu \xi^\rho) \nabla_\rho \Phi
 \ea
where in the second line we have used \eqref{dPhieq} and \eqref{2dPhieq} again. Now we use \eqref{nabla_xi} to obtain
\be
b \kappa \nabla_\mu \xi^\rho \nabla_\rho \Phi = b\kappa (\xi_\mu \eta^\rho - \eta_\mu \xi^\rho) \nabla_\rho \Phi = b\kappa (\eta^\rho \nabla_\rho \Phi) \xi_\mu
\ee
and
\be
 b( \nabla_\mu \xi^\nu)(\nabla_\nu \xi^\rho) \nabla_\rho \Phi = b ( \xi_\mu \eta^\nu - \eta_\mu \xi^\nu) \xi_\nu \eta^\rho \nabla_\rho \Phi = b \xi^\nu \eta_\nu (\eta^\rho \nabla_\rho \Phi) \xi_\mu=-b\kappa (\eta^\rho \nabla_\rho \Phi) \xi_\mu
\ee
where the final equality follows from \eqref{kappaetaxi}. Using \eqref{killing_R} we have
\be
\label{RPhi}
 b ( \xi^\nu \nabla_\mu \nabla_\nu \xi^\rho) \nabla_\rho \Phi = b R_{\rho \nu\mu\sigma} \xi^\nu \xi^\sigma \nabla^\rho \Phi
\ee
$\nabla^\mu \Phi$ is tangential to ${\cal N}$ because $\xi_\mu \nabla^\mu \Phi=0$ from \eqref{dPhieq}. Hence the RHS vanishes by \eqref{R_kappa}. 

Putting these results together we have
\be
  \nabla_\mu \left( C_{\nu\rho} \xi^\nu  \xi^\rho \right) = -2\kappa \left( a -  b\eta^\rho \nabla_\rho \Phi \right) \xi_\mu= -2A\kappa \xi_\mu = -2 \kappa C_{\mu\nu} \xi^\nu
\ee
where we used \eqref{CA} in the final two equalities. Comparing with \eqref{kappaC} we see $\kappa_C = \kappa$. 

\bibliographystyle{JHEP}
\bibliography{Remote}

\providecommand{\href}[2]{#2}\begingroup\raggedright\begin{thebibliography}{10}

\bibitem{Lovelock1971}
D.~Lovelock, {\it {The Einstein Tensor and Its Generalizations}},  {\em J.
  Math. Phys.} {\bf 12} (1971) 498--501.

\bibitem{Horndeski1974}
G.~W. Horndeski, {\it {Second-order scalar-tensor field equations in a
  four-dimensional space}},  {\em Int. J. Theor. Phys.} {\bf 10} (1974)
  363--384.

\bibitem{Weinberg:2008hq}
S.~Weinberg, {\it {Effective Field Theory for Inflation}},  {\em Phys. Rev.}
  {\bf D77} (2008) 123541, [\href{http://xxx.lanl.gov/abs/0804.4291}{{\tt
  arXiv:0804.4291}}].

\bibitem{Solomon:2017nlh}
A.~R. Solomon and M.~Trodden, {\it {Higher-derivative operators and effective
  field theory for general scalar-tensor theories}},  {\em JCAP} {\bf 02}
  (2018) 031, [\href{http://xxx.lanl.gov/abs/1709.0969}{{\tt
  arXiv:1709.0969}}].

\bibitem{Glavan:2017srd}
D.~Glavan, {\it {Perturbative reduction of derivative order in EFT}},  {\em
  JHEP} {\bf 02} (2018) 136, [\href{http://xxx.lanl.gov/abs/1710.0156}{{\tt
  arXiv:1710.0156}}].

\bibitem{Kovacs:2020pns}
A.~D. Kov\'acs and H.~S. Reall, {\it {Well-Posed Formulation of Scalar-Tensor
  Effective Field Theory}},  {\em Phys. Rev. Lett.} {\bf 124} (2020), no.~22
  221101, [\href{http://xxx.lanl.gov/abs/2003.0432}{{\tt arXiv:2003.0432}}].

\bibitem{doi:10.1063/1.522837}
G.~W. Horndeski, {\it Conservation of charge and the einstein-maxwell field
  equations},  {\em Journal of Mathematical Physics} {\bf 17} (1976), no.~11
  1980--1987.

\bibitem{Kovacs:2020ywu}
A.~D. Kov\'acs and H.~S. Reall, {\it {Well-posed formulation of Lovelock and
  Horndeski theories}},  {\em Phys. Rev. D} {\bf 101} (2020), no.~12 124003,
  [\href{http://xxx.lanl.gov/abs/2003.0839}{{\tt arXiv:2003.0839}}].

\bibitem{christodoulou2008mathematical}
D.~Christodoulou, {\em Mathematical Problems of General Relativity I}.
\newblock European Mathematical Society, 2008.

\bibitem{born2013principles}
M.~Born and E.~Wolf, {\em Principles of optics: electromagnetic theory of
  propagation, interference and diffraction of light}.
\newblock Elsevier, 2013.

\bibitem{duff1960cauchy}
G.~F. Duff, {\it The cauchy problem for elastic waves in an anisotropic
  medium},  {\em Philosophical Transactions of the Royal Society of London.
  Series A, Mathematical and Physical Sciences} {\bf 252} (1960), no.~1010
  249--273.

\bibitem{musgrave1954propagation}
M.~Musgrave, {\it On the propagation of elastic waves in aeolotropic media ii.
  media of hexagonal symmetry},  {\em Proceedings of the Royal Society of
  London. Series A. Mathematical and Physical Sciences} {\bf 226} (1954),
  no.~1166 356--366.

\bibitem{Aragone:1987jm}
C.~Aragone, {\it {Stringy characteristics of effective gravity}},  in {\em
  {SILARG 6: 6th Latin American Symposium on Relativity and Gravitation Rio de
  Janeiro, Brazil, July 13-18, 1987}}, pp.~60--69, 1987.

\bibitem{Choquet-Bruhat1988}
Y.~Choquet-Bruhat, {\it {The Cauchy problem for stringy gravity}},  {\em J.
  Math. Phys.} {\bf 29} (1988) 1891.

\bibitem{Reall2014}
H.~S. Reall, N.~Tanahashi, and B.~Way, {\it {Causality and hyperbolicity of
  Lovelock theories}},  {\em Class. Quantum Gravity} {\bf 31} (2014) 205005,
  [\href{http://xxx.lanl.gov/abs/1406.3379}{{\tt arXiv:1406.3379}}].

\bibitem{Reall2014a}
H.~S. Reall, N.~Tanahashi, and B.~Way, {\it {Shock Formation in Lovelock
  Theories}},  {\em Phys. Rev. D} {\bf 91} (2014) 044013,
  [\href{http://xxx.lanl.gov/abs/1409.3874}{{\tt arXiv:1409.3874}}].

\bibitem{Balakin:2017eur}
A.~B. Balakin and A.~E. Zayats, {\it {Non-minimal Einstein\textendash{}Maxwell
  theory: the Fresnel equation and the Petrov classification of a trace-free
  susceptibility tensor}},  {\em Class. Quant. Grav.} {\bf 35} (2018), no.~6
  065006, [\href{http://xxx.lanl.gov/abs/1710.0801}{{\tt arXiv:1710.0801}}].

\bibitem{Papallo:2017ddx}
G.~Papallo, {\it {On the hyperbolicity of the most general Horndeski theory}},
  {\em Phys. Rev.} {\bf D96} (2017), no.~12 124036,
  [\href{http://xxx.lanl.gov/abs/1710.1015}{{\tt arXiv:1710.1015}}].

\bibitem{Hajian:2020dcq}
K.~Hajian, S.~Liberati, M.~M. Sheikh-Jabbari, and M.~H. Vahidinia, {\it {On
  Black Hole Temperature in Horndeski Gravity}},  {\em Phys. Lett. B} {\bf 812}
  (2021) 136002, [\href{http://xxx.lanl.gov/abs/2005.1298}{{\tt
  arXiv:2005.1298}}].

\bibitem{Papallo:2017qvl}
G.~Papallo and H.~S. Reall, {\it {On the local well-posedness of Lovelock and
  Horndeski theories}},  {\em Phys. Rev.} {\bf D96} (2017), no.~4 044019,
  [\href{http://xxx.lanl.gov/abs/1705.0437}{{\tt arXiv:1705.0437}}].

\bibitem{Deffayet:2010qz}
C.~Deffayet, O.~Pujolas, I.~Sawicki, and A.~Vikman, {\it {Imperfect Dark Energy
  from Kinetic Gravity Braiding}},  {\em JCAP} {\bf 10} (2010) 026,
  [\href{http://xxx.lanl.gov/abs/1008.0048}{{\tt arXiv:1008.0048}}].

\bibitem{Kobayashi:2011nu}
T.~Kobayashi, M.~Yamaguchi, and J.~Yokoyama, {\it {Generalized G-inflation:
  Inflation with the most general second-order field equations}},  {\em Prog.
  Theor. Phys.} {\bf 126} (2011) 511--529,
  [\href{http://xxx.lanl.gov/abs/1105.5723}{{\tt arXiv:1105.5723}}].

\bibitem{hormander}
L.~Hormander, {\em The Analysis of Linear Partial Differential Operators II:
  Differential Operators with Constant Coefficients}.
\newblock Springer, 2004.

\bibitem{Kobayashi:2012kh}
T.~Kobayashi, H.~Motohashi, and T.~Suyama, {\it {Black hole perturbation in the
  most general scalar-tensor theory with second-order field equations I: the
  odd-parity sector}},  {\em Phys. Rev. D} {\bf 85} (2012) 084025,
  [\href{http://xxx.lanl.gov/abs/1202.4893}{{\tt arXiv:1202.4893}}]. [Erratum:
  Phys.Rev.D 96, 109903 (2017)].

\bibitem{Kobayashi:2014wsa}
T.~Kobayashi, H.~Motohashi, and T.~Suyama, {\it {Black hole perturbation in the
  most general scalar-tensor theory with second-order field equations II: the
  even-parity sector}},  {\em Phys. Rev. D} {\bf 89} (2014), no.~8 084042,
  [\href{http://xxx.lanl.gov/abs/1402.6740}{{\tt arXiv:1402.6740}}].

\bibitem{Reula:2004xd}
O.~A. Reula, {\it {Strongly hyperbolic systems in general relativity}},  {\em
  Diff. Eq.} {\bf 01} (2004) 251,
  [\href{http://xxx.lanl.gov/abs/gr-qc/0403007}{{\tt gr-qc/0403007}}].

\bibitem{courant2008methods}
R.~Courant and D.~Hilbert, {\em Methods of Mathematical Physics: Partial
  Differential Equations}.
\newblock John Wiley \& Sons, 2008.

\bibitem{atiyah1970lacunas}
M.~F. Atiyah, R.~Bott, and L.~G{\aa}rding, {\it Lacunas for hyperbolic
  differential operators with constant coefficients i},  {\em Acta mathematica}
  {\bf 124} (1970), no.~1 109--189.

\bibitem{action_principle}
D.~Christodoulou, {\em The Action Principle and Partial Differential Equations.
  (AM-146)}.
\newblock Princeton University Press, 2000.

\bibitem{Izumi:2014loa}
K.~Izumi, {\it {Causal Structures in Gauss-Bonnet gravity}},  {\em Phys. Rev.
  D} {\bf 90} (2014), no.~4 044037,
  [\href{http://xxx.lanl.gov/abs/1406.0677}{{\tt arXiv:1406.0677}}].

\bibitem{Minamitsuji:2015nca}
M.~Minamitsuji, {\it {Causal structure in the scalar\textendash{}tensor theory
  with field derivative coupling to the Einstein tensor}},  {\em Phys. Lett. B}
  {\bf 743} (2015) 272--278.

\bibitem{Tanahashi:2017kgn}
N.~Tanahashi and S.~Ohashi, {\it {Wave propagation and shock formation in the
  most general scalar\textendash{}tensor theories}},  {\em Class. Quant. Grav.}
  {\bf 34} (2017), no.~21 215003,
  [\href{http://xxx.lanl.gov/abs/1704.0275}{{\tt arXiv:1704.0275}}].

\bibitem{Shore:2007um}
G.~M. Shore, {\it {Superluminality and UV completion}},  {\em Nucl. Phys. B}
  {\bf 778} (2007) 219--258,
  [\href{http://xxx.lanl.gov/abs/hep-th/0701185}{{\tt hep-th/0701185}}].

\bibitem{Hollowood:2015elj}
T.~J. Hollowood and G.~M. Shore, {\it {Causality Violation, Gravitational
  Shockwaves and UV Completion}},  {\em JHEP} {\bf 03} (2016) 129,
  [\href{http://xxx.lanl.gov/abs/1512.0495}{{\tt arXiv:1512.0495}}].

\bibitem{Goon:2016une}
G.~Goon and K.~Hinterbichler, {\it {Superluminality, black holes and EFT}},
  {\em JHEP} {\bf 02} (2017) 134,
  [\href{http://xxx.lanl.gov/abs/1609.0072}{{\tt arXiv:1609.0072}}].

\bibitem{deRham:2020zyh}
C.~de~Rham and A.~J. Tolley, {\it {Causality in curved spacetimes: The speed of
  light and gravity}},  {\em Phys. Rev. D} {\bf 102} (2020), no.~8 084048,
  [\href{http://xxx.lanl.gov/abs/2007.0184}{{\tt arXiv:2007.0184}}].

\bibitem{Flanagan:1996gw}
E.~E. Flanagan and R.~M. Wald, {\it {Does back reaction enforce the averaged
  null energy condition in semiclassical gravity?}},  {\em Phys. Rev.} {\bf
  D54} (1996) 6233--6283, [\href{http://xxx.lanl.gov/abs/gr-qc/9602052}{{\tt
  gr-qc/9602052}}].

\bibitem{Kanti:1995vq}
P.~Kanti, N.~Mavromatos, J.~Rizos, K.~Tamvakis, and E.~Winstanley, {\it
  {Dilatonic black holes in higher curvature string gravity}},  {\em Phys. Rev.
  D} {\bf 54} (1996) 5049--5058,
  [\href{http://xxx.lanl.gov/abs/hep-th/9511071}{{\tt hep-th/9511071}}].

\bibitem{Kleihaus:2011tg}
B.~Kleihaus, J.~Kunz, and E.~Radu, {\it {Rotating Black Holes in Dilatonic
  Einstein-Gauss-Bonnet Theory}},  {\em Phys. Rev. Lett.} {\bf 106} (2011)
  151104, [\href{http://xxx.lanl.gov/abs/1101.2868}{{\tt arXiv:1101.2868}}].

\bibitem{Kleihaus:2015aje}
B.~Kleihaus, J.~Kunz, S.~Mojica, and E.~Radu, {\it {Spinning black holes in
  Einstein\textendash{}Gauss-Bonnet\textendash{}dilaton theory: Nonperturbative
  solutions}},  {\em Phys. Rev. D} {\bf 93} (2016), no.~4 044047,
  [\href{http://xxx.lanl.gov/abs/1511.0551}{{\tt arXiv:1511.0551}}].

\bibitem{East:2020hgw}
W.~E. East and J.~L. Ripley, {\it {Evolution of Einstein-scalar-Gauss-Bonnet
  gravity using a modified harmonic formulation}},
  \href{http://xxx.lanl.gov/abs/2011.0354}{{\tt arXiv:2011.0354}}.

\bibitem{Ripley:2019hxt}
J.~L. Ripley and F.~Pretorius, {\it {Hyperbolicity in Spherical Gravitational
  Collapse in a Horndeski Theory}},  {\em Phys. Rev.} {\bf D99} (2019), no.~8
  084014, [\href{http://xxx.lanl.gov/abs/1902.0146}{{\tt arXiv:1902.0146}}].

\end{thebibliography}\endgroup

\end{document}